\documentclass[aps,prd,superscriptaddress,nofootinbib,floatfix,preprint]{revtex4-1}

\usepackage[dvipdfmx]{graphicx}
\usepackage{mediabb}
\usepackage{amsmath}
\usepackage{bm}
\usepackage[small,bf]{caption}
\usepackage[subrefformat=parens]{subcaption}
\usepackage{comment}

\newcommand{\beq}{\begin{equation}}
\newcommand{\eeq}{\end{equation}}
\newcommand{\beqa}{\begin{eqnarray}}
\newcommand{\eeqa}{\end{eqnarray}}
\newcommand{\bpr}{\begin{problem}}
\newcommand{\epr}{\end{problem}}
\newcommand{\bcent}{\begin{center}}
\newcommand{\ecent}{\end{center}}
\newcommand{\bfig}{\begin{figure}}
\newcommand{\efig}{\end{figure}}
\newcommand{\bpc}{\begin{picture}}
\newcommand{\epc}{\end{picture}}
\newcommand{\barr}{\begin{array}}
\newcommand{\earr}{\end{array}}
\newcommand{\bitm}{\begin{itemize}}
\newcommand{\eitm}{\end{itemize}}
\newcommand{\bright}{\begin{flushright}}
\newcommand{\eright}{\end{flushright}}
\newcommand{\bminip}{\begin{minipage}}
\newcommand{\eminip}{\end{minipage}}
\newcommand{\btab}{\begin{tabular}}
\newcommand{\etab}{\end{tabular}}

\newcommand{\lmd}{\lambda}

\newcommand{\nnb}{\nonumber}

\newcommand{\QUP}{International Center for Quantum-field Measurement Systems for Studies of the Universe and Particles (QUP), KEK, Tsukuba, Ibaraki 305-0801, Japan}
\newcommand{\hiroshima}{Graduate School of Advanced Science and Engineering, Hiroshima University, Kagamiyama, Higashi-Hiroshima 739-8526, Japan}

\newcommand{\kyoto}{Graduate School of Science, Kyoto University, Sakyouku, Kyoto 606-8502, Japan}

\newcommand{\om}{\omega}
\newcommand{\vth}{\vartheta}

\newcommand{\mcal}[1]{\mathcal{#1}}

\newcommand{\Int}[2]{\int_{#1}^{#2}}

\newcommand{\B}{\langle}
\newcommand{\K}{\rangle}

\newcommand{\Equation}[1]{\begin{equation}#1\end{equation}}
\newcommand{\SplitEqn}[1]{\begin{equation}\begin{split}#1\end{split}\end{equation}}
\newcommand{\Align}[1]{\begin{align}#1\end{align}}

\newcommand{\AlignedEqn}[1]{\begin{equation}\begin{aligned}#1\end{aligned}\end{equation}}

\newcommand{\SubEqns}[1]{\begin{subequations}#1\end{subequations}}
\newcommand{\Cases}[1]{\begin{cases}#1\end{cases}}

\newcommand{\Exp}[1]{\exp\left[#1\right]} 
\newcommand{\ParenB}[1]{\left(#1\right)} 
\newcommand{\CurlyB}[1]{\left\{#1\right\}} 
\newcommand{\BoxB}[1]{\left[#1\right]} 

\newcommand{\vep}{\varepsilon}

\begin{document}
\title{
Opening a meV mass window for Axion-like particles
with a microwave-laser-mixed stimulated resonant photon collider
}

\author{Kensuke Homma\footnote{corresponding author}}\affiliation{\hiroshima}\affiliation{\QUP}
\author{Yuri Kirita}\affiliation{\kyoto}
\author{Takafumi Miyamaru}\affiliation{\hiroshima}
\author{Takumi Hasada}\affiliation{\hiroshima}
\author{Airi Kodama}\affiliation{\hiroshima}

\date{\today}

\begin{abstract}
We propose a microwave-laser-mixed three-beam stimulated resonant photon collider,
which enables access to axion-like particles in the meV mass range.
Collisions between a focused pulse laser beam and a focused microwave pulse beam
directly produce axion-like particles (ALPs) and another focused pulse laser beam
stimulates their decay. The expected sensitivity in the meV mass range has been evaluated.
The result shows that the sensitivity can reach the ALP-photon coupling down to 
$\mathcal{O}(10^{-13})$~GeV${}^{-1}$ with $10^6$ shots
if 10-100 TW class high-intensity lasers are properly combined with a conventional 
100 MW class S-band klystron.
This proposal paves the way for identifying the nature of ALPs as candidates for dark matter,
independent of any cosmological and astrophysical models.
\end{abstract}

\maketitle

\section{Introduction}
Dark matter (DM) is one of the most intriguing unresolved problems in modern physics.
Identifying DM in controlled experiments is thus an important subject.
On the other hand, independently from such an astronomical and cosmological issue, 
elementary particle physics requires resolution to the strong CP problem.
The topological nature of the QCD vacuum, $\theta$-vacuum, to solve the $U(1)_A$ anomaly
naturally requires CP violating nature in the QCD Lagrangian.
Nonetheless, the measured $\theta$-value in the neutron electric dipole moment 
suggests the CP conserving nature.
To this non-trivial problem, Peccei and Quinn proposed a global $U(1)_{PQ}$ symmetry~\cite{PQ}.
Through the symmetry breaking,
a counter $\theta$-value can be dynamically produced and it cancels out the finite $\theta$-value.
As a result of this symmetry breaking, axion with finite mass may appear~\cite{AXION}.
If the energy scale of the symmetry breaking is much higher than that of
the electroweak interaction, the coupling of axion to ordinary matter can be weak
and thus such an invisible axion can be a quite rational candidate for low-mass
dark matter~\cite{Preskill:1982cy,Abbott:1982af,Dine:1982ah}.
Axion cosmology may have connections to the other unresolved problems in
the standard model of elementary particle physics:
finite neutrino mass and baryon number asymmetry and also in the cosmological problems:
inflation and dark matter.
Recently, extended SU(5) grand unified theory (GUT)~\cite{SU5U1pq} and 
SO(10) GUT~\cite{SO10U1pq} by adding $U(1)_{PQ}$ to solve the strong CP problem
are discussed as an approach to access these unresolved issues within one stroke.
Cosmological observations, particularly those derived from cosmic microwave background (CMB) 
sensitive to the early-stage evolution of the Universe, present an opportune moment 
to refine our understanding by constraining the mutual parameter spaces associated 
with specific phenomena by overarching via the concept of spontaneous symmetry breaking.

There is an unexplored mass range around meV in the photon-axion
coupling predicted by the benchmark QCD axion models.
Furthermore, the unified inflaton and dark matter model 
({\it miracle})~\cite{Miracle1,Miracle2}
predicts an axion-like particle (ALP) in the photon-ALP coupling
$\mathcal{O}(10^{-10} - 10^{-13})$~$\mbox{GeV}^{-1}$
in the allowed ALP mass range $\mathcal{O}(10^{-3} - 10^2)$ eV
based on the viable parameter space consistent with the CMB observation.
Thus, opening a search window in the meV mass range would increase the
potential to discover axion and ALPs.

Historically, hallo-scope~\cite{ADMX} and helio-scope~\cite{CAST} have spearheaded 
the quest for axions and axion-like particles (ALPs), boasting the highest sensitivities 
to date. Nonetheless, while these observations may potentially detect manifestations of 
ALP decays in the future, discerning the spin and parity states of dark matter 
would pose a formidable challenge.
Therefore, as complementary observations, 
we have proposed stimulated resonant photon colliders (SRPC) using coherent electromagnetic fields
with two beams~\cite{DEptp,JHEP2020} and with three beams~\cite{3beam00}, respectively.
The method is to directly produce ALPs and simultaneously stimulate their decays 
by combining several laser fields.
Quasi-parallel SRPC with two laser beams has been
dedicated for the sub-eV axion mass window and the searches have been
actually performed~\cite{PTEP2014,PTEP2015,PTEP2020,SAPPHIRES00,SAPPHIRES01}. 
In contrast, SRPC with large incident angles of three laser beams 
has a potential to be sensitive to higher mass~\cite{3beam00,3beam01,3beam02}
compared to the two beam case.

In this paper we propose an extremely asymmetric three-beam collider by 
combining a microwave beam from a klystron and 
two short pulse laser beams in order to open a search window for meV-scale ALPs.
We then provide the expected sensitivity based on a practical set of beam parameters.

\section{Kinematical relations in extremely asymmetric stimulated three beam collider}
\begin{figure}[!h]
\begin{center}
\includegraphics[scale=0.5]{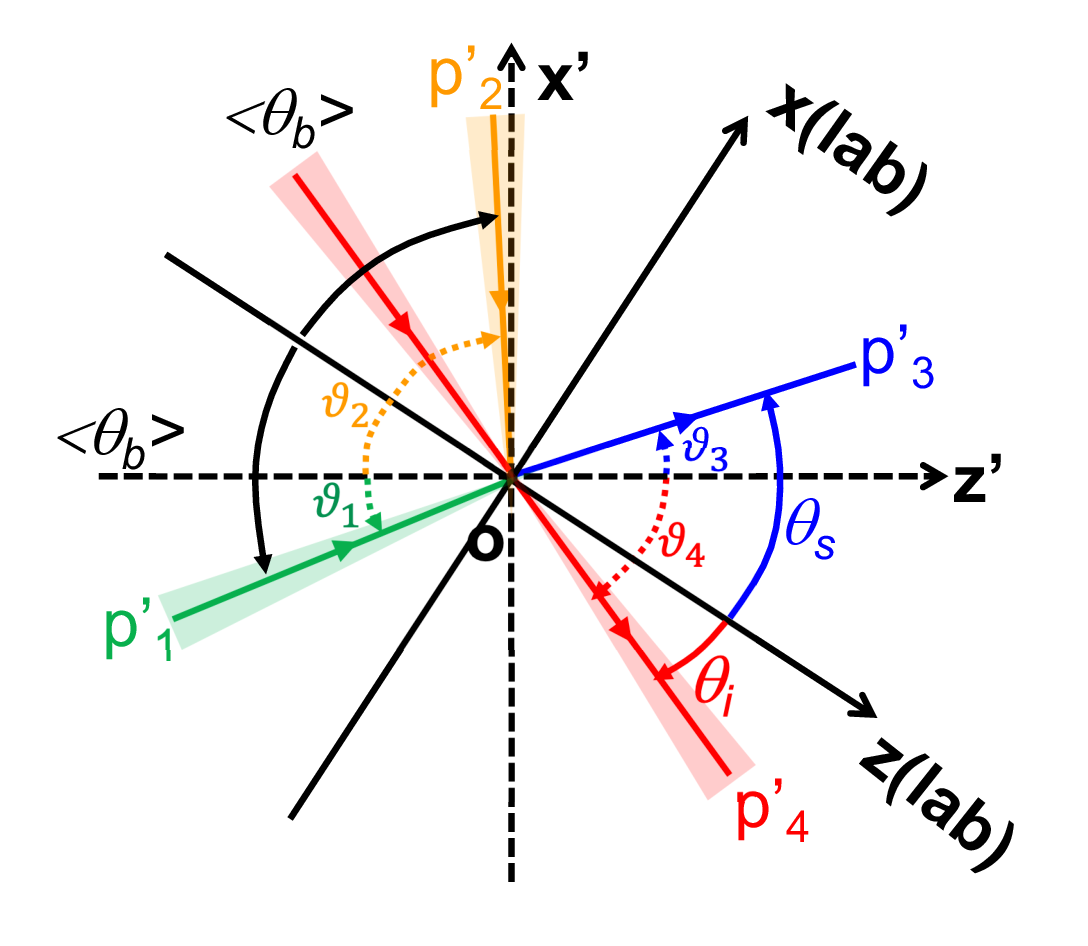}
\end{center}
\caption{
Angle relations in a three beam collider between laboratory coordinates and zero transverse
momentum coordinates with the primed symbol.
Colored cones represent focused beams at the common focal point (O):
$p^{'}_1$ and $p^{'}_2$ denote two incident photon momenta from the two creation beams
without drawing the diverging parts of the beams
while $p^{'}_4$ indicates a photon momentum from the inducing beam 
with the focusing and diverging beam. As a result of the interaction,
a signal photon, $p^{'}_3$, is emitted.
Although angles of these four photons may deviate from the individual beam axes
due to the uncertainty principles, the peculiar case where all the photons coincide
with the beam axes is on purpose displayed to introduce the angle definitions
in the two coordinate systems.
Since a target mass gives a bisecting angle $\B \vth_b \K$ between the two creation beam axes,
the bisecting axis can be a natural $z$-coordinate as the laboratory coordinates.
Further details can be found in the main text.
}
\label{Fig1}
\end{figure}

We consider photon-photon collisions to produce an ALP resonance state
between a focused short-pulse laser beam (green) and 
a focused microwave pulse beam (orange) as illustrated in Fig.\ref{Fig1}.
Simultaneously we focus another short-pulse laser beam (red) into 
the collision point in order to induce decay of the resonantly produced ALP.
In the case of collisions between tightly focused photon beams,
at around the focal region, individual photon momenta must follow
the uncertainty principle in the momentum-position relation.
In addition, in the case of short pulsed beams, 
the uncertainty on the energy-time relation must also be taken into account.
Therefore, we consider the stimulated photon-photon scattering process by including
stochastic selections of photons from the focused fields with different momenta and energies
from the central values of those in the three beams.
Suppose four-momenta $p_1$ and $p_2$ from the incident laser and microwave beams, respectively,
for the resonance creation part and $p_4$ from the inducing laser beam.
The signal photon is then defined as $p_3$ as a result of ALP decay into $p_3 + p_4$.
For a given pair of $p_1$ and $p_2$, we are allowed to arbitrarily set an axis to which incident angles
are relatively defined. 
In such arbitrary coordinates (dashed axes in Fig.\ref{Fig1}), 
with the energies of four photons $\om_j$ and scattering angles
$\vth_j$ for initial $j=1,2$ and final $j=3,4$ states, four-momenta are defined as follows:
\begin{equation}
\label{momentum}
\begin{alignedat}{10}
p_{1} & = (& \om_{1},\ & & \om_{1}\sin\vth_{1},\ & & 0,\ & & \om_{1}\cos\vth_{1})&, \\
p_{2} & = (& \om_{2},\ & & -\om_{2}\sin\vth_{2},\ & & 0,\ & & \om_{2}\cos\vth_{2})&, \\
p_{3} & = (& \om_{3},\ & & \om_{3}\sin\vth_{3},\ & & 0,\ & & \om_{3}\cos\vth_{3})&,\\
p_{4} & = (& \om_{4},\ & & -\om_{4}\sin\vth_{4},\ & & 0,\ & & \om_{4}\cos\vth_{4}).&
\end{alignedat}
\end{equation}
For later convenience, a bisecting angle $\vth_b$ is further introduced as
\beqa\label{eq_thb}
\vth_{b} \equiv \frac{\vth_1+\vth_2}{2}.
\eeqa

The energy-momentum conservation requires following relations
\begin{align}\label{eq_EnergyMomentum}
\om_1+\om_2 & = \om_3+\om_4\\ \nnb
\om_1\cos\vth_1+\om_2\cos\vth_2 & = \om_3\cos\vth_3+\om_4\cos\vth_4 \equiv \om_z\\ \nnb
\om_1\sin\vth_1-\om_2\sin\vth_2 & = \om_3\sin\vth_3-\om_4\sin\vth_4 \equiv \om_x.
\end{align}
The corresponding center-of-mass system energy, $E_{cms}$, is then expressed as
\begin{align}\label{eq_Ecms}
E_{cms} = \sqrt{(p_1+p_2)^2}=\sqrt{2\om_1\om_2\{1-\cos(\vth_1+\vth_2)\}}= 2\sqrt{\om_1\om_2}\sin\vth_b{}.
\end{align}
For a given ALP mass $m_a$, the resonance condition is defined as
\beq\label{eq_resocon}
m_a = E_{cms} = 2\sqrt{\om_1\om_2}\sin\vth_b{}.
\eeq
From Eqs.(\ref{eq_EnergyMomentum}) and (\ref{eq_resocon}),
we can derive the following relations by utilizing the fact that massless photons 
must satisfy the conditions ${p_3}^2 = {p_4}^2 =0$, that is,
\beqa\label{eq_massless}
{p_3}^2 =m^2_a - 2\om_4(\om_{1} + \om_{2} + \om_x\sin\vth_4 - \om_z\cos\vth_4) = 0 \\ \nnb
{p_4}^2 =m^2_a - 2\om_3(\om_{1} + \om_{2} - \om_x\sin\vth_3 - \om_z\cos\vth_3) = 0.
\eeqa
Among possible choices of a colliding axis, theoretically beneficial one is the axis to which 
the transverse momentum sum, $p_T$, between $p_1$ and $p_2$ vanishes, corresponding to
the case of $\om_x=0$ in Eq.(\ref{eq_EnergyMomentum}).
In the zero-$p_T$ coordinates, hereafter denoted with the prime symbol, 
the solid angle integral in the final state photons is greatly simplified
because $p_3$ and $p_4$ must also symmetrically distribute around this common axis. 
On the other hand, an experimentally convenient collision axis is the bisecting axis to which
the laser and microwave beam axes can be symmetrically aligned.
We thus introduce the laboratory coordinates $(t, x, y, z)$ so that the bisecting axis corresponds
to the $z$-axis where all the central beam axes are co-planer in the $x-z$ plane and 
the origin of time $t$ is set at the moment when the pulse peak positions
of the three beams simultaneously arrive at the common focal spot.

For the experimental setup, we introduce the central values for the photon energies
as $\om_{c_1}=\B\om_1\K$, $\om_{c_2}=\B\om_2\K$, $\om_s=\B\om_3\K$, and $\om_i=\B\om_4\K$
where $\B\quad \K$ denotes taking central values of individual distributions.
These are {\it a priori} known parameters for a given target ALP mass 
$\B m_a \K=2\sqrt{\om_{c_1}\om_{c_2}}\sin\B \vth_b\K$. 
In order to obtain the incident angle for the central axis 
of the inducing beam $\B\vth_4\K$ and the corresponding emission angle of 
the induced decay signal photon $\B\vth_3\K$, 
we apply the zero-$p_T$ coordinates to the central beam incident angles with the central
beam energies, that is, in the case of $\om_x=0$ in Eq.(\ref{eq_EnergyMomentum}). 
From $\om_1 \sin\vth_1 = \om_2 \sin\vth_2$ with $\vth_2=2\vth_b-\vth_1$ due to $\om_x=0$, 
the following relation on $\vth_1 (>0)$ is obtained
\beq
\vth_1 = \sin^{-1}\left(
\frac{\sin 2\vth_b}{\sqrt{(\frac{\om_1}{\om_2}+\cos 2\vth_b)^2+\sin^2 2\vth_b}} 
\right).
\eeq
The central value of $\om_z$, $\overline{\om_z}$, is calculable 
from Eq.(\ref{eq_massless}) with the individual central parameters.
The central angles are then expressed as
\beqa
\B\vth_3\K = \cos^{-1}\left( \frac{1}{\overline{\om_z}}(\om_{c_1} + \om_{c_2} 
- \frac{{\B m_a \K}^2}{2\om_s}) \right) \\ \nnb
\B\vth_4\K = \cos^{-1}\left( \frac{1}{\overline{\om_z}}(\om_{c_1} + \om_{c_2} 
- \frac{{\B m_a \K}^2}{2\om_i}) \right).
\eeqa
As denoted in Fig.\ref{Fig1}, the offset angle from the bisecting axis to the zero-$p_T$ axis
is $\vth_b - \vth_1$. With the offset, we eventually define the nominal emission angle of 
signal photons and the incident angle of the inducing beam in the laboratory coordinates as
\beqa
\vth_s \equiv \B\vth_3\K + (\B\vth_b\K - \B\vth_1\K) \\ \nnb
\vth_i \equiv \B\vth_4\K - (\B\vth_b\K - \B\vth_1\K).
\eeqa

\begin{figure}[!h]
\begin{center}
\includegraphics[scale=0.65]{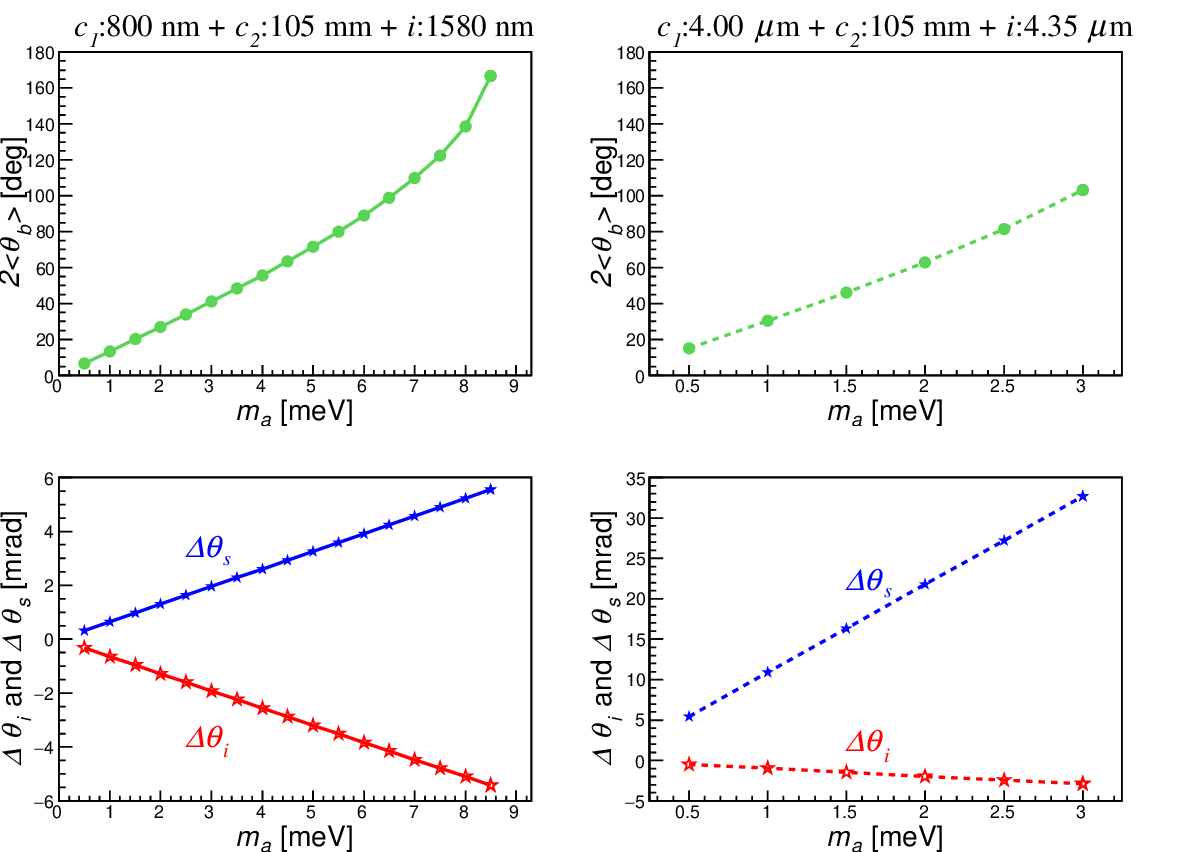}
\end{center}
\caption{
Kinematically allowed incident angles of three beams for higher (solid curves) 
and lower mass (dotted curves) options.
Left top-bottom panels are for the higher mass option consisting of
the first creation beam $c_1$: 800~nm (laser),
second creation beam $c_2$: 105~mm (S-band microwave) and
inducing beam $i$: 1580~nm (laser).
Right top-bottom panels are for the lower mass option consisting of
the first creation beam $c_1$: 4.00~$\mu$m (laser),
the second creation beam$c_2$: 105~mm (S-band microwave) and
the inducing beam $i$: 4.35~$\mu$m (laser).
Top two panels: $c_1$-beam incident angles with respect to
those of $c_2$-beam, that is, 
$2\B\vth_b\K$ as a function of targe ALP mass $\B m_a \K$.
Bottom two panels: signal emission angles and incident angles of the inducing beam
relative to incident angles of the $c_1$ beam as a function of target ALP mass $\B m_a \K$.
}
\label{Fig2}
\end{figure}

An asymmetric combination between $\om_{1}=O(1)$~eV and $\om_{2}=O(10^{-5})$~eV gives 
$E_{cms} = O(10^{-3})$~eV by adjusting the subtended angle $2\B\vth_b\K$.
In this case, however, the actual collision geometry is quite different from 
Fig.\ref{Fig1} (see Fig.\ref{Fig3})
which merely displays the defined relation between the incident angles of the beams and 
the emission angle of signals.

In this proposal we consider the following two combinations of three beams
for higher and lower mass options.
The higher mass search consists of the first creation beam $c_1$: 800~nm (laser),
the second creation beam $c_2$: 105~mm (S-band microwave) and
the inducing beam $i$: 1580~nm (laser), while
the lower mass search is the combination of the first creation beam 
$c_1$: 4.00~$\mu$m (laser), the second creation beam $c_2$: 105~mm (S-band microwave) 
and the inducing beam $i$: 4.35~$\mu$m (laser).
Figure \ref{Fig2} shows kinematically allowed incident angles of three beams 
for the higher (solid curves) and the lower mass (dotted curves) options.
The top two panels show that $c_1$-beam incident angles with respect to
$c_2$-beam, that is, 
$2\B\vth_b\K$ as a function of targe ALP mass $\B m_a \K$.
The bottom two panels display that signal emission angles and incident angles 
of the inducing beam relative to incident angles of the $c_1$ beam 
as a function of target ALP mass $\B m_a \K$.
As shown in the bottom panels, angle differences of inducing beam incident angles $\B\theta_i\K$
and those of signal photon emission angles $\B\theta_s\K$ commonly relative to $c_1$-beam incident angles
span $\cal{O}$(1-10) mrad. This indicates peculiar collision geometries
where the creation and inducing laser beams are almost parallel 
but must have slightly different intersection angles. 
Since divergence angles of typical lasers can be sub-mrad and 
the pointing stability can be guaranteed, it is feasible to precisely control 
the relative incident angles. 
The signal emission angles are also almost aligned with the $c_1$-beam incident angles. 
Because the signal photon energy $\om_s = \om_{c_1}+\om_{c_2}-\om_{i} \sim \om_{c_1}-\om_{i}$ due to
$\om_{c_1},\om_{i} \gg \om_{c_2}$ is very different from any of $\om_{c_1},\om_{c_2},\om_{i}$,
the signal photons can be separated from those of the incident laser beams
by reflecting them to another direction via a set of dichroic mirrors (see Fig.\ref{Fig3}),
which has been demonstrated in the actual experimental setups 
in~\cite{PTEP2014,PTEP2015,PTEP2020,SAPPHIRES00,SAPPHIRES01}.

\section{Conceptual design for a variable angle three beam collider}
%
\begin{figure}[]
\begin{center}
\includegraphics[scale=0.8]{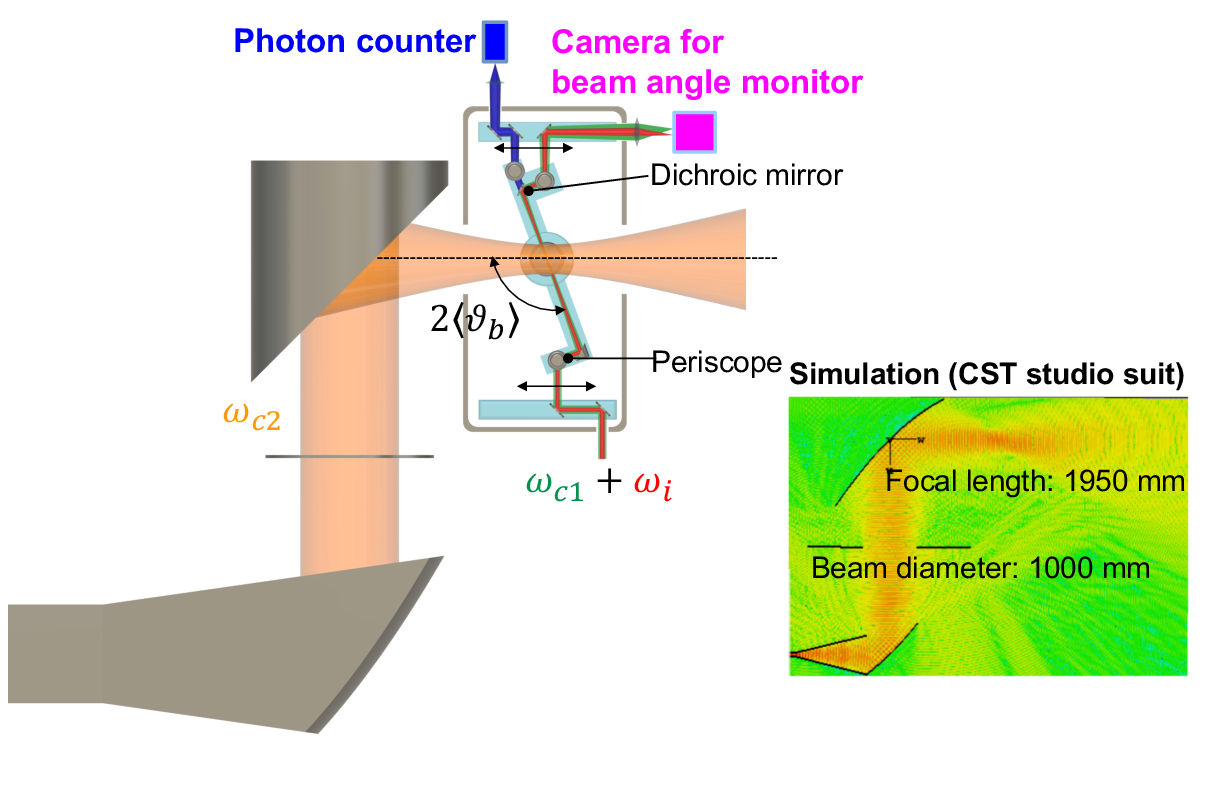}
\end{center}
\caption{
Left: schematic drawing for an experimental setup to combine laser beams and a microwave beam
for the search for meV ALPs.
The creation beam line by focusing microwaves ($c_2$) from a S-band klystron is taken to be fixed,
while the creation laser beam ($c_1$) and the inducing laser beam $(i)$ with small intersecting
angles to $c_1$ are arbitrary rotated all together on the same turning table.
As indicated in the bottom plots in Fig.\ref{Fig2}, signal photons and inducing beams
are almost aligned with the creation laser beam $c_1$. The almost co-moving laser beams are
reflected on a dichroic mirror in order to separate signal photons (blue) from 
the intense used laser beams. The incident laser beams are monitored by a camera
which measures the individual focal positions so that the slight intersecting angles
between the two lasers can be measured in order to ensure the correct kinematical relation.
The signal photons are collected and counted by a single-photon sensitive photodetector.
Right: 3D-simulation result on the S-band microwave focusing.
The result using CST Studio Suite~\cite{CST} shows that a focal distance around 2~m 
is indeed feasible if a proper horn and an aperture are equipped 
in addition to the focusing parabolic mirror.
}
\label{Fig3}
\end{figure}
Given the kinematical relations between three beams in the previous section,
we envision a searching setup for scanning the meV mass range
as depicted in Fig.\ref{Fig3}, where a microwave photon beam ($c_2$: 105 mm) from a
S-band (2.856~GHz) klystron is assumed to be focused from a fixed position,
while incident angles of a creation laser beam ($c_1$: 800~nm / 4000~nm) and 
an inducing laser beam (i: 1580~nm / 4350~nm) are variably controlled. 
The wavelength of 800~nm is available using well-known Ti:Sapphire lasers
and the high-intensity lasers are available worldwide~\cite{ELI},
the wavelength around 4~$\mu$m can be produced with
solid-state lasers based on iron-doped zinc selenide (Fe:ZnSe) 
currently attracting significant attention as efficient and powerful sources 
in the mid-infrared (MIR) spectral region~\cite{Tokita}.
The wavelengths of the individual inducing lasers can be generated via
optical parametric amplification by pumping the corresponding creation lasers
with respect to proper seed lasers. Therefore, there is some degree of choice for signal wavelengths
corresponding to $\om_{c_1} - \om_i$.
By introducing the two combinations of the laser fields, the meV mass window can be enlarged.
Focusing of a S-band microwave beam in a short distance is not a trivial issue.
We thus performed the 3D-simulation using CST Studio Suite~\cite{CST},
which is shown in the right figure. We conclude that 
a focal distance around 2~m is feasible if a proper horn and an aperture are equipped
in addition to the focusing parabolic mirror.
Since the incident angles of the two laser beams
and the signal photon emission angle are almost aligned, a rotating table accommodating
the two beams and a dichroic mirror reflecting the two beams to a different direction
from the signal's one can be implemented. 
The signal photons are assumed to be counted by a single-photon-sensitive photodevice,
while the two laser beams sent outside the chamber are further focused into a camera 
so that the focal points of the two beams can be monitored to guarantee the accurate incident 
angle relations.
According to rotation angles of the table, a sliding mirror and the top-bottom
mirrors in the periscope locally rotate so that incident points of the two lasers 
to the entrance window of the interaction chamber can remain unchanged.
 
\section{Signal yield in stimulated resonant photon scattering}
We address the following effective Lagrangian describing
the interaction of an ALP as a pseudoscalar field $\phi_a$ with two photons
\beq\label{Lagrangian}
-{\cal L} = \frac{1}{4}\frac{g}{M}F_{\mu\nu}\tilde{F}^{\mu\nu}\phi_a,
\eeq
using dimensionless coupling $g$, an energy scale for symmetry breaking $M$,
electromagnetic field strength tensor $F_{\mu\nu}$ and 
its dual $\tilde{F}^{\mu\nu}$.
The comprehensive derivation of the scattering amplitude for stimulated resonant photon
scattering applicable to the most general collision geometry is elaborated upon in
~\cite{JHEP2020,Universe00}. 
Further elucidation regarding the implementation of these formulations 
in the context of a three-beam collision scenario, featuring two equivalent incident energies 
at symmetric incident angles with an inducing beam, is presented in \cite{3beam00}. 
Hereafter, we outline the formulation by reviewing the pertinent sections 
in order to evaluate the expected sensitivity in the collision case characterized by 
two highly disparate incident energies and asymmetric incident angles with an inducing beam.

\begin{figure}[]
\begin{center}
\includegraphics[scale=0.65]{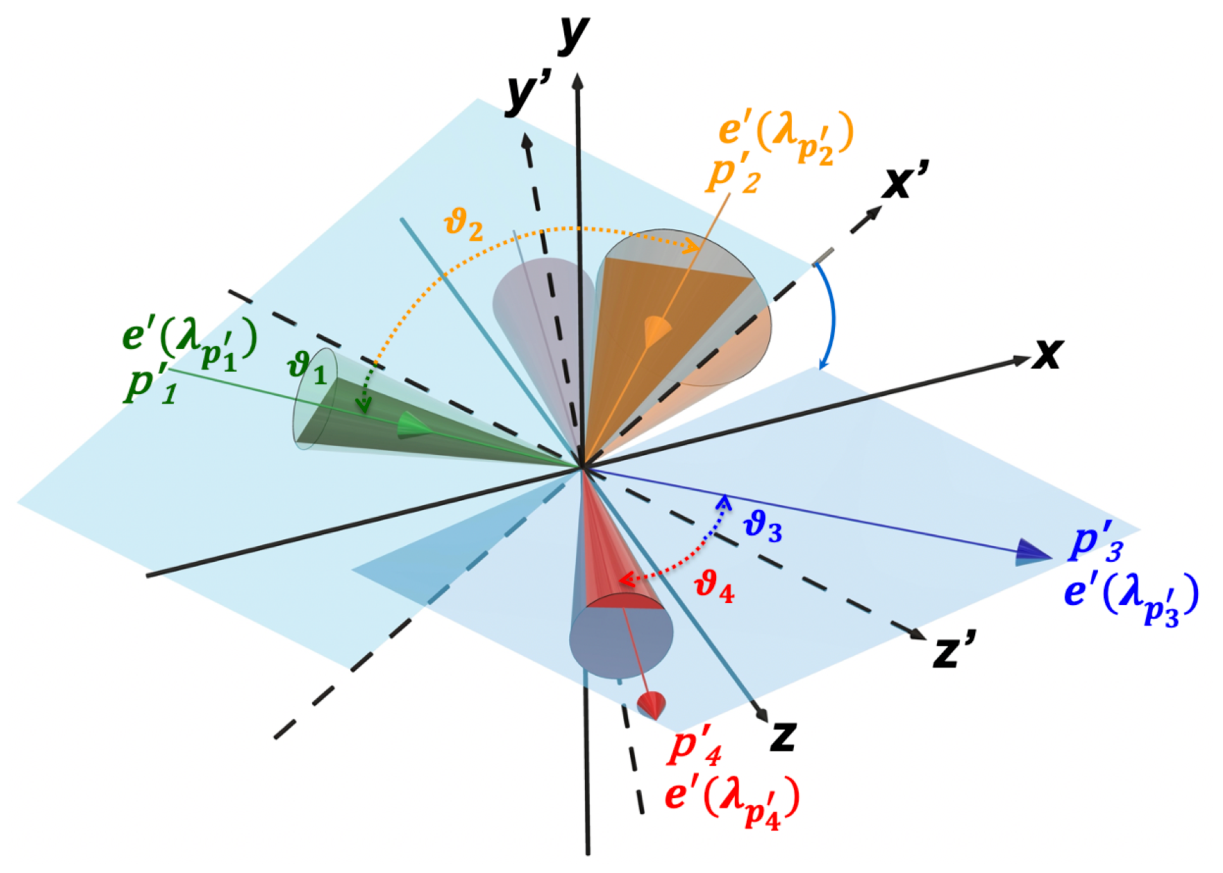}
\end{center}
\caption{Relations between laboratory coordinates and theoretical coordinates with the primed symbol.
The central axes of the three beams are physically placed on the x-z plane in laboratory coordinates
in the co-planer condition.
Theoretically a $z^{'}$-axis is calculable
so that stochastically selected two incident photons satisfying the resonance condition
have zero pair transverse momentum ($p_T$) with respect to $z^{'}$.
The Lorentz invariant scattering amplitude is calculated on the primed coordinates 
where rotation symmetries of the initial and final state reaction planes (sky blue) 
around $z^{'}$ are guaranteed.
Notations of four-momentum vectors $p'_j$ and four-polarization vectors $e'(\lambda_{p'_j})$
with polarization states $\lambda_{p'_j}$ for the initial state ($j=1,2$) and
final state ($j=3,4$) plane waves are displayed.
}
\label{Fig4}
\end{figure}

The relation between theoretical coordinates with the primed symbol and 
laboratory coordinates to which colliding beams are physically mapped is 
illustrated in Fig.\ref{Fig4}. 
By taking the momentum uncertainty, that is, the incident angle uncertainty into account,
we have to accept a situation where reaction planes between
pairs of incident photons from the two beams most likely deviate
from the co-planar plane including the three beams, that is, $x-z$ plane.
As illustrated in Fig.\ref{Fig4}, we thus introduce zero transverse momentum coordinates 
(zero-$p_T$) where the total transverse momentum of an incident pair (and outgoing pair) of photons 
becomes zero, as a basis for the calculation of the photon-photon scattering amplitude. 
We define the $z^{'}$-axis as the direction of the summed vector of the incident
pair of photons in the zero-$p_T$ coordinates by adding the transverse $x^{'}$-axis
aligned to the $p_T$ direction so that the pair vectors are contained in
the $x^{'}-z^{'}$ plane.
The zero-$p_T$ coordinates provides the axial symmetry around the $z^{'}$-axis, thus, simplifies
the calculation for the interaction rate with stimulation by the inducing beam. 
Once we calculate the rate, the positions of signal photons can be calculated
by rotating zero-$p_T$ coordinates back to the laboratory coordinates.
%
%

The Lorentz invariant scattering amplitude is computed in 
the primed coordinates, where the rotational symmetries of the initial 
and final state reaction planes around $z^{'}$ are preserved. 
Notations of four-momentum vectors $p'_j$ and four-polarization vectors $e'(\lambda_{p'_j})$
with polarization states $\lambda_{p'_j}$ for the initial-state ($j=1,2$) and
final-state ($j=3,4$) plane waves, are displayed.
The conversion between the two coordinate systems is achievable through 
a straightforward rotation $\cal{R}$, as elaborated below. 
Henceforth, unless ambiguity arises, the prime symbol associated 
with the momentum vectors will be omitted.

We commence by examining the spontaneous emission yield of the signal $p_3$, 
denoted as $\mathcal{Y}$, within the scattering process 
$p_1 + p_2 \rightarrow p_3 + p_4$, utilizing solely two incident photon beams 
characterized by normalized number densities $\rho_1$ and $\rho_2$
with average numbers of photons $N_1$ and $N_2$ for pulse 1 and 2, respectively.
The notion of "cross section" proves beneficial 
when the beams of $p_1$ and $p_2$ remain fixed. 
Nonetheless, in scenarios where the momenta of $p_1$ and $p_2$ 
exhibit significant fluctuations within the beams, 
the utility of this concept diminishes~\cite{BJ}.
Instead of "cross section", we hereby introduce the refined expression 
for "volume-wise interaction rate" denoted as $\overline{\Sigma}$~\cite{PTEP2014}, 
measured in units of length ($L$) and time ($s$), indicated as $[\quad]$:
\beqa\label{eq_Y}
{\mathcal Y} = N_1 N_2
\left(
\int dt d\bm{r} \rho_1(\bm{r},t) \rho_2(\bm{r},t)
\right)
\times \mbox{\hspace{2cm}} \\ \nnb
\left(
\int dQ W(Q)
\frac{c}{2\om_1 2\om_2} |{\mathcal M}_s(Q^{'})|^2 dL^{'}_{ips}
\right)
\mbox{\hspace{0.2cm}} \nnb\\
\equiv N_1 N_2 {\mathcal D}\left[s/L^3\right] \overline{\Sigma}\left[L^3/s\right]
\mbox{\hspace{3.3cm}}
\eeqa
with the velocity of light $c$ and the Lorentz-invariant phase space factor $dL_{ips}$
\beq\label{eqLIPS}
dL_{ips} = (2\pi)^4 \delta(p_3+p_4-p_1-p_2)
\frac{d^3p_3}{2\omega_3(2\pi)^3}\frac{d^3p_4}{2\omega_4(2\pi)^3}.
\eeq
Here, the probability density of the center-of-mass system energy, 
$W(Q)$, is incorporated to yield the average across the feasible range of $Q$.
For the incident beams $\alpha=1$ and $2$,
$W(Q)$ is defined as a function of the combinations of photon energies($\omega_{\alpha}$),
polar($\Theta_{\alpha}$) and azimuthal($\Phi_{\alpha}$) angles in laboratory coordinates,
which is denoted as
\beq\label{eq_QdQ}
Q \equiv \{\omega_{\alpha}, \Theta_{\alpha}, \Phi_{\alpha}\} \quad \mbox{and} \quad
dQ \equiv \Pi_{\alpha} d\omega_{\alpha} d\Theta_{\alpha} d\Phi_{\alpha}.
\eeq
%
The integral, weighted by $W(Q)$, embodies the resonance enhancement by 
incorporating both the off-shell component and the pole within the s-channel amplitude
corresponding to the Breit-Wigner resonance function included in $|{\mathcal M}_s(Q^{'})|^2$
with spin states $s$ defined through combinations of four-polarization vectors
$e'(\lambda_{p'_j})$ for $j=1,2,3,4$~\cite{JHEP2020}.
In primed coordinates as depicted in Fig.\ref{Fig4}, 
$Q^{'} \equiv {\omega_{\alpha}, \vth_{\alpha}, \phi_{\alpha}}$ 
represent kinematical parameters within a rotated coordinate system $Q^{'}$, constructed from 
a pair of incident wave vectors such that the transverse momentum of the pair, relative to 
a $z^{'}$-axis, is nullified. 
The conversion from $Q$ to $Q^{'}$ is thus delineated by 
rotation matrices acting upon polar and azimuthal angles: 
$\vth_{\alpha} \equiv {\cal R}_{\vth_{\alpha}}(Q)$ and 
$\phi_{\alpha} \equiv {\cal R}_{\phi_{\alpha}}(Q)$.

By introducing an inducing beam characterized by the central four-momentum $p_4$, 
possessing the normalized number density denoted as $\rho_4$ and 
the averaged photon number of photons, $N_4$, 
we broaden the scope of the \textit{spontaneous} yield 
to encompass the \textit{induced} yield, denoted as ${\cal Y}_{c+i}$, 
incorporating an expanded set of kinematical parameters as delineated below:
\beqa\label{eq_YI}
{\mathcal Y}_{c+i} = N_1 N_2 N_4
\left(
\int dt d\bm{r} \rho_1(\bm{r},t) \rho_2(\bm{r},t) \rho_4(\bm{r},t) V_4
\right) \times \mbox{\hspace{0.7cm}} \\ \nnb
\left(
\int dQ_I W(Q_I)
\frac{c}{2\om_1 2\om_2} |{\mathcal M}_s(Q^{'})|^2 dL^{'I}_{ips} \mbox{\hspace{0.1cm}}
\right)
\\ \nnb
\equiv  N_1 N_2 N_4 {\mathcal D}_{exp}\left[s/L^3\right] \overline{\Sigma}_I\left[L^3/s\right]
\mbox{\hspace{3.1cm}}
\eeqa
with
\beq\label{eq_QI}
Q_I \equiv \{Q, \omega_4, \Theta_4, \Phi_4\} \quad \mbox{and} \quad
dQ_I \equiv dQd\omega_4 d\Theta_4 d\Phi_4,
\eeq
where
the factor $\rho_4(\bm{r},t) V_4$ represents a probabilistic measure indicative of 
the extent of spatiotemporal alignment between the $p_1$ and $p_2$ beams and 
the inducing beam $p_4$, within the specified volume $V_4$ of the $p_4$ beam. 
Meanwhile, $dL^{'I}_{ips}$ expresses an inducible phase space wherein the solid angles of $p_3$ 
balance with those of $p_4$, ensuring conservation of energy-momentum within 
the distribution of the inducing beam. 
This entails the conversion of $p_4$ from the primed coordinate system to 
the corresponding laboratory coordinates, where the three beams are physically mapped
in order to estimate the effective enhancement factor due to the inducing effect.
With Gaussian distributions denoted as $G$, 
the function $W(Q_I)$ is precisely characterized as:
\begin{equation}\label{eq_WQI}
W(Q_I) \equiv \Pi_\beta G_E(\omega_{\beta}) G_p(\Theta_{\beta},\Phi_{\beta})
\end{equation}
for $\beta = 1, 2, 4$. 
Here, $G_E$, reflecting an energy spread due to the Fourier transform-limited duration of 
a short pulse, and $G_p$, representing the momentum space or equivalently the polar angle 
distribution, are introduced. These distributions are based on the properties of a focused 
coherent electromagnetic field with axial symmetric characteristics concerning 
the azimuthal angle $\Phi_{\beta}$ around the optical axis of focused beam $\beta$.

Integrating Eq.(\ref{eq_YI}) analytically is not practical.
Thus the numerical integral is performed. 
The detailed algorithm for the integral of $\overline{\Sigma}_I$ is provided in ~\cite{JHEP2020,3beam00},
while the quasi-analytic expression for the generalized density overlapping factor 
${\mathcal D}_{exp}$ in Eq.(\ref{eq_YI}) adaptable to real experimental conditions is 
provided in Appendix of this paper.

\section{Sensitivity projection}
We now evaluate the expected sensitivity with the given expression for the signal yield.
Table~\ref{Tab1} summarizes assumed beam relevant parameters for $c_1$, $c_2$, and $i$
as well as the statistical parameters.
For the set of the beam parameters $P$ in Tab. \ref{Tab1},
the number of signal photons, $N_{obs}$, in the three-beam stimulated resonant
photon scattering process is expressed as
\beq\label{Nobs}
N_{obs} = {\cal Y}_{c+i}(m_a, g/M ; P) N_{shot} \epsilon 
\eeq
as a function of ALP mass $m_a$ and coupling $g/M$
with the number of laser shots, $N_{shot}$, and
the overall efficiency for detecting $p_3$, $\epsilon$.
For a set of $m_a$ values with assumed $N_{obs}$,
a set of coupling $g/M$ can be obtained by numerically solving Eq.(\ref{Nobs}).
\begin{figure}[!h]
\begin{center}
\includegraphics[scale=0.80]{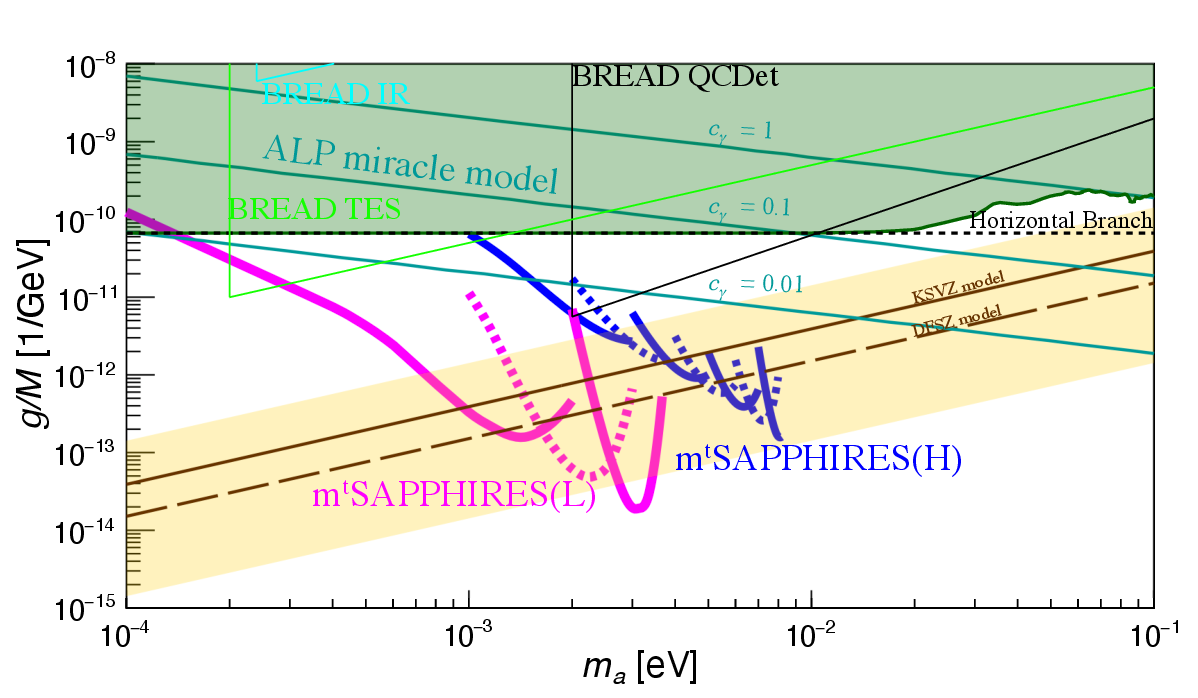}
\end{center}
\caption{
Expected sensitivities in the coupling-mass relation
for the pseudoscalar field exchange at a 95\% confidence level
by a three-beam stimulated resonant photon collider combining two laser beams and
a microwave beam from a S-band klystron.
The used beam parameters are summarized in Table \ref{Tab1}.
The thick solid/dashed curves show the expected upper limits
by the proposed laser-microwave-mixed three-beam stimulated resonant photon collider 
(m${}^t$SAPPHIRES) with the parameter set in Tab.\ref{Tab1}.
m${}^t$SAPPHIRES(H) (blue) denotes the combination between
$c_1$: 800~nm, $c_2$: S-Band, and $i$: 1580~nm,
while m${}^t$SAPPHIRES(L) (magenta) corresponds to the combination between
$c_1$: 4000~nm, $c_2$: S-Band, and $i$: 4350~nm.
Further details can be found in the main text.
}
\label{Fig5}
\end{figure}
With the parameter values in Table \ref{Tab1},
Fig.\ref{Fig5} shows the sensitivity projection in the coupling-mass relation
for the pseudoscalar field exchange at a 95\% confidence level.


These sensitivity curves are derived under the following conditions. 
In this simulated exploration, the null hypothesis assumes fluctuations 
in the count of photon-like signals, adhering to a Gaussian distribution with 
an expected value of zero for the provided total collision statistics. 
The term "photon-like signals" denotes instances where peaks resembling photons are 
enumerated through a peak-finding algorithm applied to digitized waveform data obtained from 
a photodetector~\cite{SAPPHIRES00}. 
Here, electrical fluctuations around the waveform's baseline yield 
both positive and negative counts of photon-like signals.

To reject this null hypothesis, 
a confidence level $1-\alpha$ is defined as
\beq
1-\alpha = \frac{1}{\sqrt{2\pi}\sigma}\int^{\mu+\delta}_{\mu-\delta}
e^{-(x-\mu)^2/(2\sigma^2)} dx = \mbox{erf}\left(\frac{\delta}{\sqrt 2 \sigma}\right),
\eeq
where $\mu$ represents the expected value of an estimator $x$ following the hypothesis, and
$\sigma$ denotes one standard deviation.
In this investigation, the estimator $x$ corresponds to the count of signal photons $N_S$ and
we presume the uncertainty, $\delta N_{S}$, uncorrected for detector acceptance, 
to represent one standard deviation $\sigma$ around the mean value $\mu=0$.
In order to set a confidence level of 95\%, $2 \alpha = 0.05$ with $\delta = 2.24 \sigma$ is used, 
where a one-sided upper limit by excluding above $x+\delta$~\cite{PDGstatistics} is considered.
For a given set of experimental parameters $P$ outlined in Table~\ref{Tab1},
the upper limits on the coupling--mass relation, $m_a$ vs. $g/M$,
are determined by numerically solving the following equation
\beq
N_{obs} = 2.24 \delta N_S = {\cal Y}_{c+i}(m_a, g/M ; P) N_{shots} \epsilon.
\eeq

The commonly assumed $\delta N_{S} = 50$ in Tab.\ref{Tab1}
is based on the empirical fact that there are unavoidable 
pedestal noises in typical high-intensity laser facilities~\cite{SAPPHIRES00}
even if dark current from a photo-device is negligibly small.
The signal wavelengths for the higher (H) / lower (L) mass search options:
$c_1$(800~nm), $c_2$(S-Band), and $i$(1580~nm)
/ 
$c_1$(4000~nm), $c_2$(S-Band), and $i$(4350~nm)
are expected to be $\sim 1.6$~$\mu$m (0.77 eV) / $\sim 52$~$\mu$m (24 meV), respectively.
Numerous types of photon counters for these wavelengths are now available.
For instance, commercially available photomultipliers with InP/InGaAsP photocathodes / 
superconducting optical transition-edge sensors~\cite{OptTES}
for the H-option depending on the actual choice of the inducing laser wavelength
and superconducting tunnel junction (STJ) sensors~\cite{CnuB} /
kinetic inductance detectors (KID)~\cite{MKID} / 
quantum capacitance detectors~\cite{QCDet1} for the L-option
are reasonable candidates.
From the following relation between noise equivalent power (NEP)
and dark current rate (DCR)~\cite{BREAD}
\beq\label{NEP}
\mbox{NEP} [\mbox{W}/\sqrt{\mbox{Hz}}] =
\frac{\om [\mbox{eV}]}{\epsilon [1]}\sqrt{\mbox{2DCR} [\mbox{s}^{-1}]}
\eeq
where $\om$ is signal photon energy and $\epsilon$ (assumed 10\% here) 
is overall detection efficiency with respect to the signal photon, 
DCRs can be evaluated as 
DCR(0.77 eV) = 0.329 s${}^{-1}$ / DCR(24 meV) = 338 s${}^{-1}$
for a conservative NEP = $10^{-18}$ W/$\sqrt{\mbox{Hz}}$ compared to
$\mathcal{O}(10^{-19})$ in KID~\cite{MKID} and 
$\mathcal{O}(10^{-21})$ in QCDet~\cite{QCDet1}.
For a maximum timing window due to the S-band klystron pulse duration 
$\tau_{c_2} = 1$~$\mu$s,
the accidental coincidence count defined as $2\tau_{c_2} \cdot \mbox{DCR} \cdot f \cdot T$
is expected to be 0.81 (0.77 eV) / 26 (24 meV)
with $N_{shots} = f \cdot T=10^6$ where a typical pulse repetition 
rate $f=10$~Hz for a data taking time $T=10^5$ s is assumed. 
These DCR-originating counts are sufficiently lower than $\delta N_{S} = 50$.
Therefore, this sensitivity projection offers a conservative evaluation.

The thick solid/dashed curves show the expected upper limits
by the proposed laser-microwave-mixed three-beam stimulated photon collider (m${}^t$SAPPHIRES)
with the parameter set in Tab.\ref{Tab1}.
m${}^t$SAPPHIRES(H) (blue) denotes the combination between
$c_1$: 800~nm, $c_2$: S-Band, and $i$: 1580~nm,
while m${}^t$SAPPHIRES(L) (magenta) corresponds to the combination between
$c_1$: 4000~nm, $c_2$: S-Band, and $i$: 4350~nm. 
Thanks to momentum and energy uncertainties, even for a single $\B \vth_b \K$ 
adjusted for a central mass $\B m_a \K$, the sensitive mass range can have a finite width,
which can reduce the number of different collisional geometries necessary 
to cover one order of magnitude in the mass range.
The mass scanning is assumed to be with $0.1$~meV step.
For the viewing purpose, the solid and dashed curves are alternatively depicted.
The other solid lines are sensitivity projections from the proposed
Broadband Reflector Experiment for Axion Detection (BREAD)~\cite{BREAD}
to search multiple decades of DM mass without tuning
combined with several types of photosensors:
IR Labs (blue) - cryogenic semiconducting thermistor ~\cite{IRLabs},
KID/TES (green) - kinetic inductance detectors (KID) ~\cite{MKID}
and superconducting titanium-gold transition edge sensors (TES)~\cite{TES},
QCDet (black) - quantum capacitance detectors~\cite{QCDet1,QCDet2}.
The horizontal dotted line shows the upper limit from the Horizontal Branch (HB) observation~\cite{HB}.
The green area is excluded by the helio-scope experiment CAST~\cite{CAST}.
The yellow band shows the QCD axion benchmark models
with $0.07<|E/N-1.95|<7$ where KSVZ($E/N=0$)~\cite{KSVZ} and DFSZ($E/N=8/3$)~\cite{DFSZ}
are shown with the brawn lines.
The dark cyan lines show predictions from the ALP {\it miracle} model~\cite{Miracle1,Miracle2}
with its intrinsic model parameters $c_{\gamma}=1.0, 0.1, 0.01$, respectively.
\begin{table}[h!]
\caption{Assumed experimental parameters used to numerically calculate the upper limits on 
the coupling--mass relations. The idealized ${\mathcal D}_{exp}$ without beam
drifts in Eq.(\ref{eq:[th-df]:drv_Dfactor}) with Eq.(\ref{eq:[th-df]:enum_Df_params_qps})
is applied.
}
\begin{center}
\begin{tabular}{lr}  \\ \hline
Parameter & Value \\ \hline
Centeral wavelength of creation laser ${\lambda_c}_1$ & 800~nm(H) / 4000~nm(L)\\
Relative linewidth of creation laser, $\delta{\omega_c}_1/<{\omega_c}_1>$ &  $2.0 \times 10^{-2}$\\
Duration time of creation laser, $\tau_{c_1}$ & 30 fs / 100 fs\\
Creation laser energy per $\tau_{c_1}$, $E_{c_1}$ & 1~J \\
Number of creation photons,  $N_{c_1}$ & $4.03 \times 10^{18}$(H) / $2.01 \times 10^{19}$(L) photons\\
Focal length of off-axis parabolic mirror, $f_{c_1}$ & 1.0~m\\
Beam diameter of creation laser beam, $d_{c_1}$ & 0.05~m\\
Polarization & linear (P-polarized state) \\ \hline
Central wavelength of creation laser ${\lambda_c}_2$ & 105~mm (S-band 2.856 GHz)\\
Relative linewidth of creation laser, $\delta{\omega_c}_2/<{\omega_c}_2>$ &  $1.0 \times 10^{-4}$\\
Duration time of creation laser, $\tau_{c_2}$ & 1~$\mu$s \\
Creation laser energy per $\tau_{c_2}$, $E_{c_2}$ & 100~J \\
Number of creation photons, $N_{c_2}$ & $5.28 \times 10^{25}$ photons\\
Focal length of off-axis parabolic mirror, $f_{c_2}$ & 1.95~m\\
Beam diameter of creation laser beam, $d_{c_2}$ & 1.0~m\\
Polarization & linear (S-polarized state) \\ \hline
Central wavelength of inducing laser, $\lambda_i$  & 1580~nm(H) / 4350~nm(L)\\
Relative linewidth of inducing laser, $\delta\omega_{i}/<\omega_{i}>$ &  $2.0 \times 10^{-2}$\\
Duration time of inducing laser beam, $\tau_{i}$ & 100 fs / 100 fs\\
Inducing laser energy per $\tau_{i}$, $E_{i}$ & $0.1$J \\
Number of inducing photons, $N_i$ & $7.95 \times 10^{17}$(H) / $2.19 \times 10^{18}$(L) photons\\
Focal length of off-axis parabolic mirror, $f_{i}$ & 1.0~m\\
Beam diameter of inducing laser beam, $d_{i}$ & $0.01$~m\\
Polarization & circular (left-handed state) \\ \hline
Overall detection efficiency, $\epsilon$ & 10\% \\
Number of shots, $N_{shots}$   & $10^6$ shots\\
$\delta{N}_{S}$ & 50\\
\hline
\end{tabular}
\end{center}
\label{Tab1}
\end{table}

\section{Conclusion}
Based on the concept of a three-beam stimulated resonant photon collider by
combining focused short-pulse laser beams and a focused microwave beam,
we have evaluated expected sensitivities to axion and axion-like particles 
coupling to photons.
Assuming two 10-100 TW class laser beams
and a 100 MW class microwave beam from a conventional S-band klystron,
we found that the searching method can probe ALPs in the meV mass range 
down to $g/M = {\mathcal O}(10^{-13})$~GeV${}^{-1}$.
This sensitivity can reach the unexplored domain predicted by the benchmark QCD axion models
and the unified inflaton-ALP model.
The proposed method can provide a unique opportunity to follow up search results
if axion helio- or hallo-scopes could claim any hints on the existence of ALPs 
in the current and future surveys.

\newpage
\section*{Appendix: Derivation for experimentally tuned density overlapping factor, $\mathcal{D}_{exp}$}
The $\cal{D}$-factor characterizes the extent of spacetime overlap between two creation 
pulsed beams $(j = 1, 2)$ and one inducing pulsed beam $(j = 4)$ 
when they are focused at a common focal point and their peak positions simultaneously reach 
the intersection. 
We define the spacetime intersection as the origin of the spacetime coordinates for the pulsed beams
where the focal points of the three focused pulses coincide with each other and 
their peak positions arrive at the same time.
We define the $\cal{D}$-factor with the number density distribution $\rho_j (t, \bm{r})$
normalized by the average number of photons $N_j$ for pulse beam $j$ as follows
\Equation{
        \label{eq:[th-df]:def_Dfactor}
        \mcal{D}
                \equiv \Int{- z_{4,R} / c_0}{0} dt
                        \Int{- \infty}{\infty} d^3 \bm{r}
                        \ParenB{
                                \prod_{j = 1, 2, 4} \rho_j (t, \bm{r})
                        }
                        V_4,
}
where $z_{4,R}$ is the Rayleigh length of a focused inducing pulse
and $V_{4}$ is the volume of that.
While interactions with ALPs theoretically occur even at lower pulse intensities 
before reaching the common focal point (i.e., the origin), 
we adopt a conservative finite range from $-z_{4,R} / c_0$ to $0$ for the time integral. 
This choice reflects that the predominant fraction of interactions takes place 
as the pulses traverse the interval of the Rayleigh length just before reaching the beam waist.
Subsequently, we derive an experimentally tuned $\mathcal{D}$-factor, ${\cal{D}}_{exp}$, 
designed to estimate systematic uncertainties in the spacetime overlapping factor caused 
by beam drifts during data collection. 
For ${\cal{D}}_{exp}$, new parameters for drift in space and time for beam $j$, 
denoted as $\bm{r}_{j,0}$ and $t_{j,0}$, respectively, are introduced relative to the spacetime origin.

%
%

When Gaussian pulse beam $j$ propagating along the $z$-axis is focused at the origin of spacetime,
the normalized photon number density distribution $\rho_j$ is expressed as follows~\cite{PTEP2017}
\Equation{
        \label{eq:[th-df]:def_rho}
        \rho_j (t, \bm{r}\,;\ \lambda_j, \tau_j, d_j, f_j)
                =       \ParenB{
                                \frac{2}{\pi}
                        }^\frac{3}{2}
                        \frac{1}{w_j^2 (c_0 t) c_0 \tau_j}
                        \Exp{
                                - 2 \frac{x^2 + y^2}{w_j^2 (c_0 t)}
                        }
                        \Exp{
                                - 2
                                \ParenB{
                                        \frac{z - c_0 t}{c_0 \tau_j}
                                }^2
                        },
}
where focusing angle $\vth_{j,0}$, beam waist $w_{j,0}$, Rayleigh length $z_{j,R}$,
and spot size $w_j (c_0 t)$~\cite{Yariv} at $z = c_0 t$ with the velocity of light $c_0$
are respectively defined as
\Equation{
        \label{eq:[th-df]:enum_fundamental_params}
        \vth_{j,0}
                = \tan^{-1}\ParenB{\frac{d_j}{2 f_j}},
        \quad
        w_{j,0}
                = \frac{\lambda_j}{\pi\vth_{j, 0}},
        \quad
        z_{j,R}
                = \frac{\pi w_{j, 0}^2}{\lambda_j},
        \quad
        w_j (c_0 t)
                = w_{j,0} \sqrt{1 + \frac{(c_0 t)^2}{z_{j,R}^2}}.
}

\begin{figure}[!hbt]
        \centering
        \includegraphics[keepaspectratio, scale=0.8]{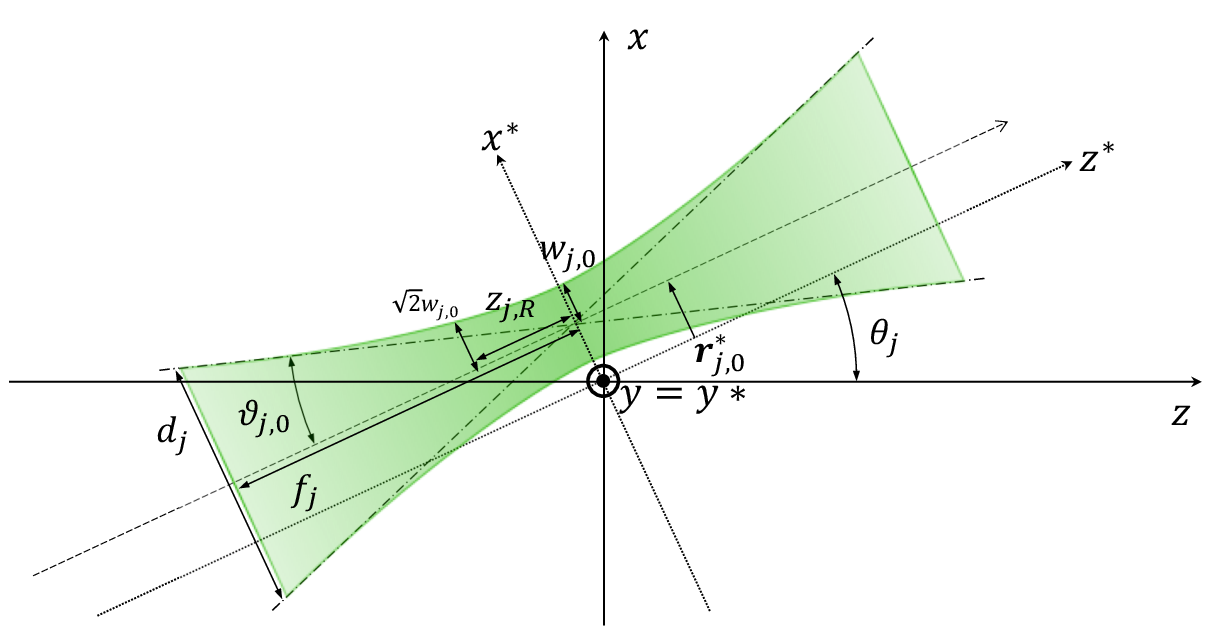}
        \caption{
                Drifted geometry and parameters for a focused laser pulse with incident angle $\theta_j$.
                The laser-beam coordinate system rotated by $\theta_j$ is denoted
                with the asterisk symbol.
                A laser pulse propagates along the dashed arrow parallel to the $z^*$-axis
                with a displacement of $\bm{r}^*_{j,0}$ from that in the $x^*-y^*$ plane.
                The focusing angle between the dashed arrow and the dash-dotted line
                based on geometric optics
                is expressed as $\vth_{j,0} = \tan^{-1} \ParenB{d_j / 2 f_j}$
                using diameter $d_j$ and focal length $f_j$ of beam $j$.
                The beam radius at the focal point is referred to as beam waist $w_{j,0}$, 
                and the beam radius along the $z^*$-axis is expressed as $w_j$.
                When $w_j = \sqrt{2} w_{j,0}$, the distance from the focal point
                corresponds to Rayleigh length $z_{j,R}$.
        }
        \label{fig:[th-df]:def_laser_params}
\end{figure}

The basic parameters characterizing geometrical properties of focused laser pulse $j$
are wavelength $\lmd_j$, pulse duration $\tau_j$, diameter $d_j$ and focal length $f_j$.
In addition, as illustrated in Fig. \ref{fig:[th-df]:def_laser_params},
we introduce parameters to reflect the experimental reality concerning
spatial rotation and spatio-temporal translation of a laser pulse:
incident angle $\theta_j$ from the $z$-axis in the $z-x$ plane,
arrival time deviation $t_{j,0}$ at the focal point,
and spatial drift $\bm{r}^*_{j,0}$ in the $x^*-y^*$ plane perpendicular to the direction of 
laser pulse propagation along $z^{*}$.
These parameters can generalize collisional geometry and thus allow to 
analyze the three beam pulse overlap factor in a real experimental condition.
Local time $t_j$ for pulse beam $j$ with time offset $t_{j,0}$ 
is introduced with respect to global time $t$ whose origin is set at the moment 
when the peak position of pulse $j$ arrives at the common focal point
\Equation{
        \label{eq:[th-df]:def_time_trf}
        t \rightarrow t_j = t - t_{j,0}.
}
Hereinafter, we refer to coordinate systems where $z^*$-axes are the individual
directions of propagating laser pulses as laser-beam coordinate systems 
to distinguish them from non-asterisked laboratory coordinates.
Since three laser pulses are generally used in SRPC, 
the three beam coordinate systems are individually considered.
A point $\bm{r}$ in the laboratory coordinate system is expressed
with the point $\bm{r}^*$ in the laser-beam coordinate system as follows
\Equation{
        \label{eq:[th-df]:def_coord_trf}
        \bm{r}
                = \mcal{R}_j (\bm{r}^* + \bm{r}^*_{j,0}),
}
where $\mcal{R}_j$ is a rotation matrix in the $z-x$ plane counterclockwise
through incident angle $\theta_j$ with respect to the positive $y$-axis.
Equation \eqref{eq:[th-df]:def_coord_trf} can also be converted to the expression 
for $\bm{r}^*$ and we write down the elements of the rotation matrix and vectors as follows
\Equation{
        \label{eq:[th-df]:wrdn_coord_trf}
        \bm{r}^*
                = \mcal{R}_j^{-1} \bm{r} - \bm{r}^*_{j,0}
                = \begin{pmatrix}
                                \cos \theta_j & 0 & - \sin \theta_j \\
                                0                                                       & 1 & 0 \\
                                \sin \theta_j & 0 & \cos \theta_j \\
                        \end{pmatrix}
                        \begin{pmatrix}
                                x \\
                                y \\
                                z \\
                        \end{pmatrix}
                        -
                        \begin{pmatrix}
                                x^*_{j,0} \\
                                y^*_{j,0} \\
                                0 \\
                        \end{pmatrix}.
}
\\
The normalized number density distribution $\rho_j$ is then generalized
by substituting Eq.\eqref{eq:[th-df]:def_time_trf} and Eq.\eqref{eq:[th-df]:wrdn_coord_trf} 
to $\rho_j (t_j, \bm{r}^*)$ as follows
\SplitEqn{
        \label{eq:[th-df]:def_gnrd_rho}
        & \rho_j (t, \bm{r}\,;\ \lambda_j, \tau_j, d_j, f_j, \theta_j, t_{j,0}, \bm{r}^*_{j,0})
        \\
        & \quad =
                \ParenB{\frac{2}{\pi}}^\frac{3}{2}
                \frac{1}{w_j^2 (c_0 t_j) c_0 \tau_j}
                \Exp{
                        - 2
                        \frac{
                                (x \cos \theta_j - z \sin \theta_j - x^*_{j,0})^2
                                +
                                (y - y^*_{j,0})^2
                        }{
                                w_j^2 (c_0 t_j)
                        }
                }
                \\
        & \qquad
                \Exp{
                        - 2
                        \ParenB{
                                \frac{x \sin \theta_j + z \cos \theta_j - c_0 t_j}{c_0 \tau_j}
                        }^2
                }.
}
Hereinafter we abbreviate
\Equation{
        \label{eq:[th-df]:def_abbr_spotsize}
        w_j = w_j(c_0 t_j).
}
The volume of an inducing pulse $V_4$ for the normalization is a quantity
which is independent of the space-time coordinates
and obtained by spatial integration of the squared field strength
of the inducing laser pulse \cite{Yariv,PTEP2017},
\Equation{
        \label{eq:[th-df]:def_inducible_vol}
        V_4
                =       \ParenB{\frac{\pi}{2}}^\frac{3}{2}
                        w_{4,0}^2 c_0 \tau_4.
}
Substituting the space-time distributions $\rho_j (t, \bm{r})$
into Eq.\eqref{eq:[th-df]:def_Dfactor}, we obtain
\SplitEqn{
        \label{eq:[th-df]:wrdn_Dfactor}
        \mcal{D}_{exp}
                & =     
                        V_4
                        \ParenB{
                                \frac{2}{\pi}
                        }^\frac{9}{2}
                        \Int{- z_{4,R} / c_0}{0} dt
                        \ParenB{\prod_j \frac{1}{w_j^2 c_0 \tau_j}}
                        \Int{-\infty}{\infty} dx
                        \Int{-\infty}{\infty} dy
                        \Int{-\infty}{\infty} dz
                        \\
                & \quad 
                        \Exp{   
                                - 2 \sum_j
                                \CurlyB{
                                        \ParenB{
                                                \frac{x \cos \theta_j - z \sin \theta_j - x^*_{j,0}}{w_j}
                                        }^2
                                        +
                                        \ParenB{
                                                \frac{y - y^*_{j,0}}{w_j}
                                        }^2
                                        +
                                        \ParenB{
                                                \frac{x \sin \theta_j + z \cos \theta_j - c_0 t_j}{c_0 \tau_j}
                                        }^2
                                }
                        }.
}

In the following, first, all the spatial integrations will be performed. 
The operation that is most frequently repeated is
a re-square-completion of several square-completed quadratic functions.
The sum over $j$ of square-completed quadratic functions expressed in one variable $x$
can be transformable into a single square-completed quadratic function as follows
\Equation{
        \label{eq:[th-df]:dfm_gaus_qdrsum}
        \sum_j a_j (x - b_j)^2
                =       \ParenB{\sum_j a_j}
                        \CurlyB{
                                x - \frac{\sum_j a_j b_j}{\sum_j a_j}
                        }^2
                        +
                        \frac{\sum_{j, k, j < k} a_j a_k (b_j - b_k)^2}{\sum_j a_j}
}
The summation contained in the numerator of the last term in the RHS
of Eq.\eqref{eq:[th-df]:dfm_gaus_qdrsum} is over $(j, k) = (1, 2), (1, 4), (2, 4)$.
Since the summed function which consists of squares of anti-symmetric elements
holds $(j, k) = (k, j)$ with respect to the index combinations
as in the LHS of Eq.\eqref{eq:[th-df]:def_cycsum2},
this function can be regarded as a summation over $j, k$ following the cyclic order of $\{1, 2, 4\}$
\Align{
        \label{eq:[th-df]:def_cycsum2}
        \sum_{j, k, j < k} (c_j d_k - c_k d_j)^2
                & =     \sum_{j, k = (1, 2), (1, 4), (2, 4)} (c_j d_k - c_k d_j)^2 \nnb \\
                & =     \sum_{j, k = (2, 4), (4, 1), (1, 2)} (c_j d_k - c_k d_j)^2 \nnb \\
                & =     \sum_j \ParenB{\sum_{k,l} \vep_{jkl} c_k d_l}^2.
}
Hereafter, the cyclic summation over $j$ and $k$ is abbreviated
using the exceptional three-dimensional anti-symmetric symbol $\vep_{jkl}$,
which takes the sign of cyclic and anti-cyclic permutations of $\{1, 2, 4\}$, as follows
\Equation{
        \label{eq:[th-df]:def_vep3d}
        \vep_{jkl}
                =       \Cases{
                                1  & (j, k, l) = (1, 2, 4), (2, 4, 1), (4, 1, 2) \\
                                -1 & (j, k, l) = (1, 4, 2), (4, 2, 1), (2, 1, 4) \\
                                0        & \text{otherwise}.
                        }
}

First, the $y$-integration in Eq.(\ref{eq:[th-df]:wrdn_Dfactor}) is performed.
In the exponent of the integrand, the $y$-dependent terms are only the second terms.
The sum over $j$ of these terms can be combined into a single quadratic function
by using Eq.\eqref{eq:[th-df]:dfm_gaus_qdrsum} and Eq.\eqref{eq:[th-df]:def_cycsum2},
\Equation{
        \label{eq:[th-df]:pfm_y_sqcpl}
        \sum_j \frac{1}{w_j^2} \ParenB{y - y^*_{j,0}}^2
                =       \ParenB{\sum_j w_j^{-2}}
                        \ParenB{
                                y - \frac{\sum_j w_j^{-2} y^*_{j,0}}{\sum_j w_j^{-2}}
                        }^2
                        + \sum_j \eta_j^2,
}
where $\eta_j$ is defined as
\Equation{
        \label{eq:[th-df]:def_eta}
        \eta_j
                =       \frac{1}{\sqrt{\sum_m w_m^{-2}}}
                        \sum_{k,l}
                        \vep_{jkl}
                        w_k^{-1} w_l^{-1}
                        y^*_{k,0}.
}
The $y$-integral of the exponential term with Eq.\eqref{eq:[th-df]:pfm_y_sqcpl} can be immediately
perfomed as follows
\Equation{
        \label{eq:[th-df]:pfm_y_int}
        \Int{-\infty}{\infty}dy
        \Exp{
                - 2 \sum_j \frac{1}{w_j^2} \ParenB{y - y^*_{j,0}}^2
        }
                =       \sqrt{\frac{\pi}{2}}
                        \ParenB{
                                \sum_j w_j^{-2}
                        }^{-\frac{1}{2}}
                        \Exp{
                                - 2 \sum_j \eta_j^2
                        }.
}

Next, we turn our attention to the $x$-integral 
in Eq.\eqref{eq:[th-df]:wrdn_Dfactor}. All the terms that persist 
after the $y$-integral exhibit the negative quadratic behavior with respect to $x$ within 
the exponential function, and the range of $x$-integral spans from $-\infty$ to $\infty$. 
Consequently, we can consolidate the coefficients of all the terms dependent on $x$ 
by completing the square for $x$, facilitating the subsequent integral with respect to $x$.
We bundle the first and third terms in the curly bracket in Eq.\eqref{eq:[th-df]:wrdn_Dfactor}
with individual coefficients of $x$, respectively, as follows
\SplitEqn{
        \label{eq:[th-df]:dfm_expterms}
        &       \ParenB{
                        \frac{x \cos \theta_j - z \sin \theta_j - x^*_{j,0}}{w_j}
                }^2
                +
                \ParenB{
                        \frac{x \sin \theta_j + z \cos \theta_j - c_0 t_j}{c_0 \tau_j}
                }^2
                \\
        & =
                \ParenB{
                        \frac{\cos \theta_j}{w_j}
                }^2
                \ParenB{
                        x - z \tan \theta_j - x^*_{j,0} \sec \theta_j
                }^2
                +
                \ParenB{
                        \frac{\sin \theta_j}{c_0 \tau_j}
                }^2
                \ParenB{
                        x + z \cot \theta_j - c_0 t_j \csc \theta_j
                }^2.
}
Hereinafter, we define $\alpha_j$ and $\beta_j$ as
\Equation{
        \label{eq:[th-df]:def_alphabeta}
        \alpha_j
                =       \frac{1}{w_j^2} \cos \theta_j
        , \qquad
        \beta_j
                =       \frac{1}{c_0^2 \tau_j^2} \sin \theta_j.
}
Furthermore, we temporarily replace for the sake of simplification,
\Equation{
        \label{eq:[th-df]:def_ABCD}
        A_j = \alpha_j \cos \theta_j
        , \quad
        B_j = z \tan \theta_j + x^*_{j,0} \sec \theta_j
        , \quad
        C_j = \beta_j \sin \theta_j
        , \quad
        D_j = z \cot \theta_j - c_0 t_j \csc \theta_j.
}
Equation \eqref{eq:[th-df]:dfm_expterms} can be transformable into the re-completed  
square for $x$ as follows,
\Align{
        \label{eq:[th-df]:pfm_x_sqcpl}
        &       A_j \ParenB{x - B_j}^2 + C_j \ParenB{x + D_j}^2
                \nnb \\
        & =
                \ParenB{A_j + C_j}
                \CurlyB{
                        x - \frac{1}{A_j + C_j} \ParenB{A_j B_j - C_j D_j}
                }^2
                +
                \frac{1}{A_j + C_j}
                \frac{1}{\ParenB{ w_j c_0 \tau_j }^2}
                \ParenB{
                        z + x^*_{j,0} \sin \theta_j - c_0 t_j \cos \theta_j
                }^2
                \nnb \\
        & =
                \delta_j
                \ParenB{x - E_j}^2
                +
                \frac{1}{\delta_j}
                \frac{1}{\ParenB{w_j c_0 \tau_j}^2}
                \ParenB{z + F_j}^2,
}
where $\delta_j$, $E_j$ and $F_j$ are defined as
\Align{
        \label{eq:[th-df]:def_delta}
        \delta_j
                &       =       A_j + C_j
                        =       \alpha_j \cos \theta_j + \beta_j \sin \theta_j,
                        \\
        \label{eq:[th-df]:def_E}
        E_j
                & =
                        \frac{1}{\delta_j}
                        \ParenB{
                                A_j B_j - C_j D_j
                        }
                        \nnb \\
                & =
                        \frac{1}{\delta_j}
                        \ParenB{
                                \ParenB{
                                        \alpha_j \sin \theta_j - \beta_j \cos \theta_j
                                } z
                                +
                                \alpha_j x^*_{j,0}
                                +
                                \beta_j c_0 t_j
                        },
                        \\
        \label{eq:[th-df]:def_F}
        F_j & = x^*_{j,0} \sin \theta_j - c_0 t_j \cos \theta_j.
}
$\delta_j$ is always positive for Eq.\eqref{eq:[th-df]:def_alphabeta}.
The sum over $j$ of the last term in Eq.\eqref{eq:[th-df]:pfm_x_sqcpl} is calculated
by using Eq.\eqref{eq:[th-df]:dfm_gaus_qdrsum} and Eq.\eqref{eq:[th-df]:def_cycsum2} as follows
\SplitEqn{
        \label{eq:[th-df]:pfm_x_sqcpl2}
        &       \sum_j
                \CurlyB{
                        \delta_j \ParenB{x - E_j}^2
                        +
                        \frac{1}{\delta_j}
                        \frac{1}{\ParenB{w_j c_0 \tau_j}^2}
                        \ParenB{z + F_j}^2
                }
                \\
        & =
                \ParenB{\sum_j \delta_j}
                \ParenB{
                        x -  \frac{\sum_j \delta_j E_j}{\sum_j \delta_j}
                }^2
                +
                \frac{
                        \sum_j
                        \ParenB{
                                \sum_{k,l} \vep_{jkl} \sqrt{\delta_k \delta_l} E_k
                        }^2
                }{
                        \sum_j \delta_j
                }
                +
                \sum_j
                \frac{1}{\delta_j}
                \frac{1}{\ParenB{w_j c_0 \tau_j}^2}
                \ParenB{z + F_j}^2.
}
We substitute Eq.\eqref{eq:[th-df]:pfm_x_sqcpl2} into the exponential term
in Eq.\eqref{eq:[th-df]:wrdn_Dfactor} and calculate the $x$-integration as follows
\Align{
        \label{eq:[th-df]:pfm_x_int}
        &       \Int{-\infty}{\infty}dx
                \Exp{
                        - 2
                        \sum_j
                        \CurlyB{
                                \delta_j \ParenB{x - E_j}^2
                                +
                                \frac{1}{\delta_j}
                                \frac{1}{\ParenB{w_j c_0 \tau_j}^2}
                                \ParenB{z + F_j}^2
                        }
                }
                \nnb \\
        & =
                \sqrt{\frac{\pi}{2}}
                \ParenB{
                        \sum_j \delta_j
                }^{-\frac{1}{2}}
                \Exp{
                        - 2
                        \CurlyB{
                                \frac{
                                        \sum_j
                                        \ParenB{
                                                \sum_{k,l} \vep_{jkl} \sqrt{\delta_k \delta_l} E_k
                                        }^2
                                }{
                                        \sum_j \delta_j
                                }
                                +
                                \sum_j
                                \frac{1}{\delta_j}
                                \frac{1}{\ParenB{w_j c_0 \tau_j}^2}
                                \ParenB{z + F_j}^2
                        }
                }.
}

We then perform the $z$-integral. The term that depends on $z$ is only the exponential term
in the RHS of Eq.\eqref{eq:[th-df]:pfm_x_int}.
The $z$-dependent terms, like the $x$-dependent terms,
are all negative quadratic within the exponential,
and the integral range is from $-\infty$ to $\infty$.
Therefore, the plan is that all the terms are summarized by the square completion for $z$
and the gaussian integral of them is performed.
The first term in the curly bracket in Eq.\eqref{eq:[th-df]:pfm_x_int} is square-completed at first.
We transform $\sum_{k,l} \vep_{jkl} \sqrt{\delta_k \delta_l} E_k$ to the expression 
that denotes $z$ explicitly as follows
\Align{
        \label{eq:[th-df]:dfm_Eterm}
        \sum_{k,l} \vep_{jkl} \sqrt{\delta_k \delta_l} E_k
                & =
                        \sqrt{\sum_m \delta_m}
                        \CurlyB{
                                \frac{1}{\sqrt{\sum_m \delta_m}}
                                \sum_{k,l} \vep_{jkl} \sqrt{\frac{\delta_l}{\delta_k}}
                                \ParenB{
                                        \alpha_k \sin \theta_k - \beta_k \cos \theta_k
                                } z
                        \right.
                        \\
                & \qquad
                        \left.
                                +
                                \frac{1}{\sqrt{\sum_m \delta_m}}
                                \sum_{k,l} \vep_{jkl} \sqrt{\frac{\delta_l}{\delta_k}}
                                \ParenB{
                                        \alpha_k x^*_{k,0} + \beta_k c_0 t_k
                                }
                        }
                        \nnb \\
                & =
                        \sqrt{\sum_m \delta_m}
                        \ParenB{\mu_j z + \xi_j},
}
where $\mu_j$ and $\xi_j$ are defined as
\SplitEqn{
        \label{eq:[th-df]:def_muxi}
        \mu_j
                & =
                        \frac{1}{\sqrt{\sum_m \delta_m}}
                        \sum_{k,l} \vep_{jkl} \sqrt{\frac{\delta_l}{\delta_k}}
                        \ParenB{
                                \alpha_k \sin \theta_k - \beta_k \cos \theta_k
                        },
        \qquad
        \\[4pt]
        \xi_j
                & =
                        \frac{1}{\sqrt{\sum_m \delta_m}}
                        \sum_{k,l} \vep_{jkl} \sqrt{\frac{\delta_l}{\delta_k}}
                        \ParenB{
                                \alpha_k x^*_{k, 0} + \beta_k c_0 t_k
                        }.
}
The first term in the curly bracket in the exponential function in Eq.\eqref{eq:[th-df]:pfm_x_int} 
can be square-completed using Eq.\eqref{eq:[th-df]:dfm_gaus_qdrsum} and Eq.\eqref{eq:[th-df]:def_cycsum2}
\Align{
        \label{eq:[th-df]:pfm_z_sqcpl}
        \frac{1}{\sum_j \delta_j}
        \sum_j \ParenB{\sum_{k,l} \vep_{jkl} \sqrt{\delta_k \delta_l} E_k}^2
                & =
                        \sum_j \mu_j^2
                        \ParenB{
                                z + \frac{\xi_j}{\mu_j}
                        }^2
                        \nnb \\
                & =
                        \ParenB{\sum_j \mu_j^2}
                        \ParenB{
                                z + \frac{\sum_j \mu_j \xi_j}{\sum_j \mu_j^2}
                        }^2
                        +
                        \frac{
                                \sum_j \ParenB{\sum_{k,l} \vep_{jkl} \mu_j \xi_k}^2
                        }{
                                \sum_j \mu_j^2
                        }.
}
The second term is also the sum of quadratic functions for $z$ distinguished by the subscript $j$,
which is then square-completed to the single quadratic function with respect to $z$
using Eq.\eqref{eq:[th-df]:dfm_gaus_qdrsum} as follows
\Align{
        \label{eq:[th-df]:pfm_z_sqcpl2}
        \sum_j
        \frac{1}{\delta_j}
        \frac{1}{\ParenB{w_j c_0 \tau_j}^2}
        \ParenB{z + F_j}^2
                & =
                        \sum_j
                        \ParenB{
                                \frac{1}{w_j c_0 \tau_j \sqrt{\delta_j}} z
                                +
                                \frac{1}{w_j c_0 \tau_j \sqrt{\delta_j}}
                                \ParenB{
                                        x^*_{j,0} \sin \theta_j - c_0 t_j \cos \theta_j
                                }
                        }^2
                        \nnb \\[12pt]
                & =
                        \sum_j \ParenB{\nu_j z + \zeta_j}^2
                        \nnb \\
                & =
                        \ParenB{\sum_j \nu_j^2}
                        \ParenB{
                                z + \frac{\sum_j \nu_j \zeta_j}{\sum_j \nu_j^2}
                        }^2
                        +
                        \frac{
                                \sum_j
                                \ParenB{\sum_{k,l} \vep_{jkl} \nu_k \zeta_l}^2
                        }{
                                \sum_j \nu_j^2
                        },
}
where
\SplitEqn{
        \label{eq:[th-df]:def_nuzeta}
        \nu_j
                =       \frac{1}{w_j c_0 \tau_j \sqrt{\delta_j}}
        , \qquad
        \zeta_j
                =       \frac{1}{w_j c_0 \tau_j \sqrt{\delta_j}}
                        \ParenB{
                                x^*_{j,0} \sin \theta_j - c_0 t_j \cos \theta_j
                        }.
}
The terms depending on $z$ enumerated in Eq.\eqref{eq:[th-df]:pfm_x_int}
leave the two terms after the respective formula transformations:
the individual first terms in the second lines of the RHS
in Eq.\eqref{eq:[th-df]:pfm_z_sqcpl} and Eq.\eqref{eq:[th-df]:pfm_z_sqcpl2}.
They are further square-completed as follows
\SplitEqn{
        \label{eq:[th-df]:pfm_z_sqcpl3}
        &       \ParenB{\sum_j \mu_j^2}
                \ParenB{
                        z + \frac{\sum_j \mu_j \xi_j}{\sum_j \mu_j^2}
                }^2
                +
                \ParenB{\sum_j \nu_j^2}
                \ParenB{
                        z + \frac{\sum_j \nu_j \zeta_j}{\sum_j \nu_j^2}
                }^2
                \\
        & =
                \ParenB{
                        \sum_j \ParenB{\mu_j^2 + \nu_j^2}
                }
                \ParenB{
                        z +
                        \frac{
                                \sum_j \ParenB{\mu_j \xi_j + \nu_j \zeta_j}
                        }{
                                \sum_j \ParenB{\mu_j^2 + \nu_j^2}
                        }
                }^2
                +
                \frac{
                        \ParenB{\sum_j \mu_j^2}
                        \ParenB{\sum_j \nu_j^2}
                }{
                        \sum_j \mu_j^2 + \sum_j \nu_j^2
                }
                \ParenB{
                        \frac{\sum_j \mu_j \xi_j}{\sum_j \mu_j^2}
                        -
                        \frac{\sum_j \nu_j \zeta_j}{\sum_j \nu_j^2}
                }^2.
}
Then, the first term in Eq.\eqref{eq:[th-df]:pfm_z_sqcpl3} can be $z$-integrated
\SplitEqn{
        \label{eq:[th-df]:pfm_z_int}
        &       \Int{-\infty}{\infty}dz
                \Exp{
                        - 2
                        \ParenB{
                                \sum_j \ParenB{\mu_j^2 + \nu_j^2}
                        }
                        \ParenB{
                                z +
                                \frac{
                                        \sum_j \mu_j \xi_j + \sum_j \nu_j \zeta_j
                                }{
                                        \sum_j \ParenB{\mu_j^2 + \nu_j^2}
                                }
                        }^2
                }
                \\
        & \quad =
                \sqrt{\frac{\pi}{2}}
                \ParenB{
                        \sum_j \ParenB{\mu_j^2 + \nu_j^2}
                }^{-\frac{1}{2}}.
}

All the spatial integrations are now completed.
Based on Eq.\eqref{eq:[th-df]:wrdn_Dfactor},
the terms obtained by integrations are simply enumerated in the product form
as shown in Eq.\eqref{eq:[th-df]:sbst_all_int}.
In total, there are two major terms by the spatial integrations:
(I) the result from the $y$-integration in Eq.\eqref{eq:[th-df]:pfm_y_int},
(II) the results from the $x$- and $z$-integrations
in Eq.\eqref{eq:[th-df]:pfm_x_int} and Eq.\eqref{eq:[th-df]:pfm_z_int}
and three residual terms due to re-square-completions with respect to $z$
in Eq.\eqref{eq:[th-df]:pfm_z_sqcpl}, Eq.\eqref{eq:[th-df]:pfm_z_sqcpl2}
and Eq.\eqref{eq:[th-df]:pfm_z_sqcpl3}.
\SplitEqn{
        \label{eq:[th-df]:sbst_all_int}
        \mcal{D}_{exp}
                & =
                        V_4
                        \ParenB{\frac{2}{\pi}}^\frac{9}{2}
                        \Int{- z_{4,R} / c_0}{0} dt
                        \ParenB{\prod_j \frac{1}{w_j^2 c_0 \tau_j}}
                        \times
                        \sqrt{\frac{\pi}{2}}
                        \ParenB{\sum_j w_j^{-2}}^{-\frac{1}{2}}
                        \Exp{- 2 \sum_j \eta_j^2}
                        \\
                & \qquad
                        \times
                        \sqrt{\frac{\pi}{2}}
                        \ParenB{
                                \sum_j \delta_j
                        }^{-\frac{1}{2}}
                        \sqrt{\frac{\pi}{2}}
                        \ParenB{
                                \sum_j \ParenB{\mu_j^2 + \nu_j^2}
                        }^{-\frac{1}{2}}
                        \Exp{
                                - 2
                                \frac{
                                        \sum_j \ParenB{\sum_{k, l} \vep_{jkl} \mu_k \xi_l}^2
                                }{
                                        \sum_j \mu_j^2
                                }
                        }
                        \\
                & \qquad \quad
                        \Exp{
                                - 2
                                \frac{
                                        \sum_j \ParenB{\sum_{k, l} \vep_{jkl} \nu_k \zeta_l}^2
                                }{
                                        \sum_j \nu_j^2
                                }
                        }
                        \Exp{
                                - 2
                                \frac{
                                        \ParenB{\sum_j \mu_j^2}
                                        \ParenB{\sum_j \nu_j^2}
                                }{
                                        \sum_j \ParenB{\mu_j^2 + \nu_j^2}
                                }
                                \ParenB{
                                        \frac{\sum_j \mu_j \xi_j}{\sum_j \mu_j^2}
                                        -
                                        \frac{\sum_j \nu_j \zeta_j}{\sum_j \nu_j^2}
                                }^2
                        }.
}
The numerators of the two of the last three exponential terms can be simplified as
\SubEqns{
        \label{eq:[th-df]:smpl_expterms}
        \Align{
                \label{eq:[th-df]:smpl_exp1}
                \sum_j \ParenB{\sum_{k,l} \vep_{jkl} \mu_k \xi_l}^2
                & =
                        \ParenB{\sum_j \mu_j^2}
                        \ParenB{\sum_j \xi_j^2}
                        -
                        \ParenB{\sum_j \mu_j \xi_j}^2,
                        \\
                \label{eq:[th-df]:smpl_exp2}
                \sum_j \ParenB{\sum_{k,l} \vep_{jkl} \nu_k \zeta_l}^2
                & =
                        \ParenB{\sum_j \nu_j^2}
                        \ParenB{\sum_j \zeta_j^2}
                        -
                        \ParenB{\sum_j \nu_j \zeta_j}^2.
        }
}

Finally, the experimental $\cal{D}$-factor is derived as
\SplitEqn{
        \label{eq:[th-df]:drv_Dfactor}
        \mcal{D}_{exp}
                & =
                        \ParenB{\frac{2}{\pi}}^\frac{3}{2}
                        w_{4,0}^2 c_0 \tau_4
                        \Int{- z_{4,\mathit{R}} / c_0}{0} dt
                        \frac{
                                1
                        }{
                                \sqrt{
                                        \ParenB{\sum_j w_j^{-2}}
                                        \ParenB{\sum_j \delta_j}
                                        \ParenB{\sum_j \ParenB{\mu_j^2 + \nu_j^2}}
                                }
                        }
                        \\
                & \qquad
                        \ParenB{
                                \prod_j \frac{1}{w_j^2 c_0 \tau_j}
                        }
                        \Exp{
                                - 2
                                \CurlyB{
                                        \sum_j \ParenB{\xi_j^2 + \eta_j^2 + \zeta_j^2}
                                        -
                                        \frac{
                                                \ParenB{
                                                        \sum_j \ParenB{\mu_j \xi_j + \nu_j \zeta_j}
                                                }^2
                                        }{
                                                \sum_j \ParenB{\mu_j^2 + \nu_j^2}
                                        }
                                }
                        }.
}
Since there is the complicated time-dependent exponential term,
the analytical time integration over the finite integral range is not practical.
The parameters used in Eq.\eqref{eq:[th-df]:drv_Dfactor}
(
$\alpha_j, \beta_j$ in Eq.\eqref{eq:[th-df]:def_alphabeta};
$\delta_j$ in Eq.\eqref{eq:[th-df]:def_delta};
$\mu_j, \xi_j$ in Eq.\eqref{eq:[th-df]:def_muxi};
$\nu_j, \zeta_j$ in Eq.\eqref{eq:[th-df]:def_nuzeta};
$\eta_j$ in Eq.\eqref{eq:[th-df]:def_eta}
)
are summarized in the following set of parameters
\AlignedEqn{
        \label{eq:[th-df]:enum_Df_params}
        \alpha_j
                & =
                        \frac{1}{w_j^2} \cos \theta_j,
        \qquad
        \beta_j
                = \frac{1}{c_0^2 \tau_j^2} \sin \theta_j,
        \qquad
        \delta_j
                = \alpha_j \cos \theta_j + \beta_j \sin \theta_j,
                        \hspace{120pt}
        \\[6pt]
        \mu_j
                & =
                        \frac{1}{\sqrt{\sum_m \delta_m}}
                        \sum_{k,l} \vep_{jkl}
                        \sqrt{\frac{\delta_l}{\delta_k}}
                        \ParenB{
                                \alpha_k \sin \theta_k - \beta_k \cos \theta_k
                        },
        \qquad
        \nu_j
                = \frac{1}{w_j c_0 \tau_j \sqrt{\delta_j}},
        \\[4pt]
        \xi_j
                & =
                        \frac{1}{\sqrt{\sum_m \delta_m}}
                        \sum_{k,l} \vep_{jkl}
                        \sqrt{\frac{\delta_l}{\delta_k}}
                        \ParenB{
                                \alpha_k x^*_{k, 0} + \beta_k c_0 t_k
                        },
        \qquad
        \eta_j
                =       \frac{1}{\sqrt{\sum_m w_m^{-2}}}
                        \sum_{k,l} \vep_{jkl} w_k^{-1} w_l^{-1} y^*_{k,0},
        \\[8pt]
        \zeta_j
                & =
                        \frac{1}{w_j c_0 \tau_j \sqrt{\delta_j}}
                        \ParenB{
                                x^*_{j,0} \sin \theta_j - c_0 t_j \cos \theta_j
                        }.
}
The parameters $\xi_j, \eta_j, \zeta_j$ are caused by the drifts of propagation directions 
of individual pulses in the $x^*-y^*$ plane, $x^*_{j, 0}, y^*_{j, 0}$.
Especially, both $\xi_j$ and $\zeta_j$ explicitly include $x^*_{j, 0}$ and $c_0 t_j$
since $x$ and $z$ components are correlated by rotations of laser pulse propagation directions
in the $z-x$ plane.

When the focused spacetime points of individual pulses reach the oring,
the temporal drift becomes $t_{j,0} = 0$ and the spatial drifts
$x^*_{j, 0} = y^*_{j, 0} = 0$ are satisfied. 
The parameters for ${\cal{D}}_{exp}$ in such an ideal case are thus summarized as
\AlignedEqn{
        \label{eq:[th-df]:enum_Df_params_qps}
        \alpha_j
                & =
                        \frac{1}{w_j^2} \cos \theta_j,
        \qquad
        \beta_j
                = \frac{1}{c_0^2 \tau_j^2} \sin \theta_j,
        \qquad
        \delta_j
                = \alpha_j \cos \theta_j + \beta_j \sin \theta_j,
                        \hspace{120pt}
        \\[6pt]
        \mu_j
                & =
                        \frac{1}{\sqrt{\sum_m \delta_m}}
                        \sum_{k,l} \vep_{jkl}
                        \sqrt{\frac{\delta_l}{\delta_k}}
                        \ParenB{
                                \alpha_k \sin \theta_k - \beta_k \cos \theta_k
                        },
        \qquad
        \nu_j
                = \frac{1}{w_j c_0 \tau_j \sqrt{\delta_j}},
        \\[4pt]
        \xi_j
                & =
                        \frac{1}{\sqrt{\sum_m \delta_m}}
                        \sum_{k,l} \vep_{jkl}
                        \sqrt{\frac{\delta_l}{\delta_k}}
                        \beta_k c_0 t,
        \qquad
        \eta_j =        0,
        \qquad
        \zeta_j
                =       - \frac{1}{w_j c_0 \tau_j \sqrt{\delta_j}}
                        c_0 t \cos \theta_j.
}

In order to cross-check the complicated formulae for ${\cal{D}}_{exp}$,
we compare the $\cal{D}$-factor, ${\cal{D}}_{c+i}$, applied to 
a quasi-parallel collision system (QPS)
consisting of two beams, creation beam ($c$) and inducing beam ($i$),
which is the simplest and practical collisional system dedicated 
for SRPC~\cite{SAPPHIRES00}. In QPS,
creation photons are arbitrarily selected from a single common pulse laser
resulting in an ALP creation.
Since creation and inducing laser pulses co-axially propagate in QPS
and are focused by an off-axis parabolic mirror, these incident angles are $\theta_j = 0$.
The corresponding parameters in Eq.\eqref{eq:[th-df]:enum_Df_params} are then as follows
\AlignedEqn{
        \label{eq:[th-df]:enum_QDf_params}
        \alpha_j
                & = \frac{1}{w_j^2},
        \qquad
        \beta_j = 0,
        \qquad
        \delta_j = \frac{1}{w_j^2},
        \qquad
        \mu_j = 0,
        \qquad
        \nu_j = \frac{1}{c_0 \tau_j},
        \hspace{132pt}
        \\[4pt]
        \xi_j
                & =
                        \frac{1}{\sqrt{\sum_m w_m^{-2}}}
                        \sum_{k,l} \vep_{jkl} w_k^{-1} w_l^{-1} x^*_{k, 0},
        \quad
        \eta_j
                =       \frac{1}{\sqrt{\sum_m w_m^{-2}}}
                        \sum_{k,l} \vep_{jkl} w_k^{-1} w_l^{-1} y^*_{k,0},
        \quad
        \zeta_j = - \frac{t_j}{\tau_j}.
}
For the sake of clarity, we introduce notations for QPS as follows
$w_c \equiv w_1 = w_2$ and $w_i \equiv w_4$, $\tau_c \equiv \tau_1 = \tau_2$ 
and $\tau_i \equiv \tau_4$.
The $\cal{D}$-factor in QPS, $\mcal{D}_{c+i}^Q$, is transformed
by substituting the parameters in Eq.\eqref{eq:[th-df]:enum_QDf_params}
into Eq.\eqref{eq:[th-df]:drv_Dfactor} as follows
\Align{
        \label{eq:[th-df]:drv_QPSDfactor}
        \mcal{D}_{c+i}^Q
                & =
                        \ParenB{\frac{2}{\pi}}^\frac{3}{2}
                        \frac{1}{c_0}
                        \frac{\tau_i}{\tau_c}
                        \frac{1}{\sqrt{\tau_c^2 + 2 \tau_i^2}}
                        w_{i,0}^2
                        \Int{- z_{i,R}/c_0}{0} dt
                        \frac{1}{w_c^2 \ParenB{w_c^2 + 2 w_i^2}}
                        \Exp{
                                - 2 \sum_j \ParenB{\xi_j^2 + \eta_j^2}
                        }.
}
The exponential term in Eq.\eqref{eq:[th-df]:drv_QPSDfactor} expresses
the deviations between the profiles of two laser pulses at the focal point as
\Equation{
        \label{eq:[th-df]:dfm_QDf_devterm}
        \sum_j \ParenB{\xi_j^2 + \eta_j^2}
                =       \frac{2}{w_c^2 + 2 w_i^2}
                        \CurlyB{
                                \ParenB{x^*_{c, 0} - x^*_{i, 0}}^2
                                +
                                \ParenB{y^*_{c, 0} - y^*_{i, 0}}^2
                        }.
}

As the ideal situation in QPS, 
we consider the case where time drift $t_{j,0}$ and spatial drift $x^*_{j, 0}, y^*_{j, 0}$ are absent.
Since Eq.\eqref{eq:[th-df]:dfm_QDf_devterm} becomes null, Eq.\eqref{eq:[th-df]:drv_QPSDfactor} is expressed as
\Equation{
        \label{eq:[th-df]:smpl_QPSDfactor}
        \mcal{D}_{c+i}^Q
                =       \ParenB{\frac{2}{\pi}}^\frac{3}{2}
                        \frac{1}{c_0}
                        \frac{\tau_i}{\tau_c}
                        \frac{1}{\sqrt{\tau_c^2 + 2 \tau_i^2}}
                        w_{i,0}^2
                        \Int{- z_{i,R}/c_0}{0} dt
                        \frac{1}{w_c^2 (c_0 t) \ParenB{w_c^2 (c_0 t) + 2 w_i^2 (c_0 t)}},
}
where
\Equation{
        \label{eq:[th-df]:dcmp_QDf_fracterm}
        \frac{1}{w_c^2 (c_0 t) \ParenB{w_c^2 (c_0 t) + 2 w_i^2 (c_0 t)}}
                =       \frac{1}{2\ParenB{w_{c,0}^2 \vth_{i,0}^2 -w_{i,0}^2 \vth_{c,0}^2}}
                        \BoxB{
                                \frac{\vth_{c,0}^2}{w_c^2 (c_0 t)}
                                -
                                \frac{\vth_{c,0}^2 + 2\vth_{i,0}^2}{w_c^2 (c_0 t) + 2 w_i^2 (c_0 t)}
                        }.
}
Eventually, $\mcal{D}_{c+i}^Q$ is thus simplified as
\Equation{
        \label{eq:[th-df]:pfm_QDf_timeint}
        \mcal{D}_{c+i}^Q
                =       \sqrt{\frac{2}{\pi^3}}
                        \frac{1}{c_0^2}
                        \frac{\tau_i}{\tau_c}
                        \frac{1}{\sqrt{\tau_c^2 + 2 \tau_i^2}}
                        \frac{
                                1
                        }{
                                \vth_{c,0}^2
                                \ParenB{1 - \frac{z_{c,R}^2}{z_{i,R}^2}}
                        }
                        \BoxB{
                                \frac{1}{z_{c,R}}
                                \tan^{-1} \ParenB{\frac{z_{i,R}}{z_{c,R}}}
                                -
                                \frac{1}{Z_{ci}}
                                \tan^{-1} \ParenB{\frac{z_{i,R}}{Z_{ci}}}
                        }
}
with
\Equation{
        \label{eq:[th-df]:def_Zci}
        Z_{ci}
                \equiv
                        \sqrt{
                                \frac{
                                        w_{c,0}^2 + 2 w_{i,0}^2
                                }{
                                        \vth_{c,0}^2 + 2 \vth_{i,0}^2
                                }
                        }.
}
This expression exactly coincides with that in the published paper~\cite{SAPPHIRES00}, 
which is derived starting from the idealized QPS case.

\section*{Acknowledgments}
K. Homma acknowledges the support of the Collaborative Research
Program of the Institute for Chemical Research (ICR) at Kyoto University
(Grant No. 2024--95)
and Grants-in-Aid for Scientific Research No. 21H04474 from the Ministry of Education,
Culture, Sports, Science and Technology (MEXT) of Japan.
%
We thank Shigeki Tokita (ICR, Kyoto Univ.) for providing information on 
the mid-infrared laser development,
Masanori Wakasugi (ICR) and Tetsuo Abe (KEK) for discussions on available klystron sources, and
Kaori Hattori (AIST/QUP, KEK) and Taiji Fukuda (AIST) for information on the optical TES sensor.
And also we thank Yuji Takeuchi (Tsukuba Univ.) and Takashi Iida (Tsukuba Univ.) 
for their explanations on the STJ sensor.

\end{document}